\documentclass[dvips,a4paper,12pt]{article}
\usepackage{amsmath}
\usepackage{graphicx}
\usepackage{authblk}
\usepackage[bookmarks=true,colorlinks=true,linkcolor=blue,citecolor=magenta]{hyperref}
\usepackage[numbers,sort&compress]{natbib}
\newcommand{\be}{\begin{equation}}
\newcommand{\en}{\end{equation}}
\newcommand{\bea}{\begin{eqnarray}}
\newcommand{\ena}{\end{eqnarray}}
\newcommand{\lbl}[1]{\label{eq:#1}}
\newcommand{\lbltab}[1]{\label{tab:#1}}
\newcommand{\lblfig}[1]{\label{fig:#1}}
\newcommand{\lblsec}[1]{\label{sec:#1}}
\newcommand{\rf}[1]{(\ref{eq:#1})}
\newcommand{\Table}[1]{\ref{tab:#1}}
\newcommand{\fig}[1]{\ref{fig:#1}}
\newcommand{\sect}[1]{\ref{sec:#1}}
\newcommand{\braque}[1]{{\langle #1 \rangle}}
\newcommand{\abstwo}[1]{{\left\vert #1 \right\vert^2}}

\newcommand{\bc}{\begin{center}}
\newcommand{\ec}{\end{center}}
\newcommand{\bt}{\begin{tabular}}
\newcommand{\et}{\end{tabular}}
\newcommand{\bg}{\begin{minipage}}
\newcommand{\eg}{\end{minipage}}
\newcommand{\ba}{\begin{array}}
\newcommand{\ea}{\end{array}}
\newcommand{\gapprox}{%
\mathrel{%
\setbox0=\hbox{$>$}\raise0.6ex\copy0\kern-\wd0\lower0.65ex\hbox{$\sim$}}}
\newcommand{\lapprox}{%
\mathrel{%
\setbox0=\hbox{$<$}\raise0.6ex\copy0\kern-\wd0\lower0.65ex\hbox{$\sim$}}}
\newcommand{\inleft}{%
\mathrel{%
\setbox0=\hbox{$<$}\copy0\kern-0.5\wd0\lower1.1\ht0\hbox{$\scriptstyle{in}$}}}
\newcommand{\inright}{%
\mathrel{%
\setbox0=\hbox{$>$}\copy0\kern-0.5\wd0\lower1.1\ht0\hbox{$\scriptstyle{in}$}}}
\newcommand{\outleft}{%
\mathrel{%
\setbox0=\hbox{$<$}\copy0\kern-0.5\wd0\lower1.1\ht0\hbox{$\scriptstyle{out}$}}}
\newcommand{\outright}{%
\mathrel{%
\setbox0=\hbox{$>$}\copy0\kern-0.5\wd0\lower1.1\ht0\hbox{$\scriptstyle{out}$}}}

\newcommand{\eslash}{%
\mathrel{%
\setbox0=\hbox{$\slash$}\copy0\kern-\wd0\hbox{$\epsilon_2$}}}

\newcommand{\eslashstar}{%
\mathrel{%
\setbox0=\hbox{$\slash$}\copy0\kern-\wd0\hbox{$\epsilon^*_2$}}}

\newcommand{\kslashone}{%
\mathrel{%
\setbox0=\hbox{$\slash$}\copy0\kern-\wd0\hbox{$k_1$}}}
\newcommand{\kslashtwo}{%
\mathrel{%
\setbox0=\hbox{$\slash$}\copy0\kern-\wd0\hbox{$k_2$}}}
\def\XXint#1#2#3{{\setbox0=\hbox{$#1{#2#3}{\int}$}
     \vcenter{\hbox{$#2#3$}}\kern-.5\wd0}}

\newcommand{\e}{{\rm e}}
\newcommand{\im}{{\rm Im\,}}
\newcommand{\re}{{\rm Re\,}}
\newcommand{\disc}{{\rm disc}}
\newcommand{\Kbar}{\bar{K}}
\newcommand{\piz}{{\pi^0}}
\newcommand{\pip}{{\pi^+}}
\newcommand{\mpi }{m_\pi}
\newcommand{\mpid}{m_\pi^2}
\newcommand{\mtaud}{m_\tau^2}
\newcommand{\mvd}{m_V^2}

\newcommand{\momegd}{m_\omega^2}

\newcommand{\fpid}{F_\pi^2}

\newcommand{\mkd}{m_K^2}

\newcommand{\mmd}{m_-^2}
\newcommand{\mpd}{m_+^2}

\newcommand{\qdd}{q^2}
\newcommand{\med}{m_e^2}

\newcommand{\betapi}{\sigma_\pi}

\newcommand{\betaK}{\sigma_K}
\newcommand\TT{\rule{0pt}{2.5ex}}         %top strut// espaces dans les tables
\newcommand\B{\rule[-0.5ex]{0pt}{0pt}}  %bottom strut
\title{Unified dispersive approach to real and virtual photon-photon
  scattering at low energy}

\author{B. Moussallam}

\affil{Groupe de Physique Th\'eorique,
    IPN, Universit\'e Paris-Sud 11, F-91406 Orsay, France}

\begin{document}

\date{\today}

\maketitle
%\tableofcontents

\begin{abstract}
Previous representations of pion pair production amplitudes by two real
photons at low energy, which combine dispersion theoretical
constraints with elastic unitarity, chiral symmetry and soft photon
constraints are generalized to the case where one photon is virtual.
The constructed amplitudes display explicitly the dependence on the
$\pi\pi$ phase-shifts, on pion form factors and on pion
polarizabilities. They  apply both for space-like and
time-like virtualities despite the apparent overlap of the left and
right-hand cuts, by implementing a definition of resonance exchange
amplitudes complying with analyticity and consistent limiting
prescriptions for the energy variables.
Applications are made to the pion generalized polarizabilies, to
vector meson radiative decays, and to the $\sigma\gamma$
electromagnetic form factor.  
Finally, an evaluation of the contribution of $\gamma\pi\pi$
states in the hadronic vacuum polarization to the muon $g-2$ is given,
which should be less model dependent than previous estimates.
\end{abstract}

\section{Introduction}
A precise knowledge of the amplitudes for producing a small number of
pions from a pair of real or virtual photons is needed for a reliable
evaluation of the hadronic light-by-light contribution to the muon $g-2$. 
Recently, the contribution which involves the $\pi\pi$  intermediate
states was evaluated to NNLO in the chiral
expansion~\cite{Engel:2012xb}. The convergence of the chiral expansion
is somewhat slowed down, in this context, by the strong attraction of
the $\pi\pi$ pair in the isoscalar $S$-wave and work is in
progress~\cite{privatemartin} aimed at going beyond the
chiral regime, by making use of relations, via unitarity, 
with the amplitudes $\gamma\gamma^*\to\pi\pi$,
$\gamma^*\gamma^*\to\pi\pi$. 

From an experimental point of view, such amplitudes are measurable at
$e^+ e^-$ colliders from $e^+ e^-\to e^+ e^- \pi\pi$. If no tagging is
performed, the cross-section is dominated by the scattering of two
quasi-real photons. By tagging one  of the final state leptons
(see~\cite{Carimalo:1979hq} for a detailed discussion of this situation) it
becomes possible to access the scattering amplitude 
between one real and one virtual  photon
in the kinematical region where the virtuality is negative.
Alternatively, from $e^+ e^-$ annihilation, one may generate the
amplitude $\gamma^*\to \gamma\pi\pi$ via final-state radiation (FSR),
which probes positive virtualities, $q^2 > 4\mpid$. In the case of
a pair of neutral pions, FSR is the only possible mechanism. Experimental
measurements of the cross-section for  $e^+ e^-\to\gamma\pi^0\pi^0$
have recently been performed by  several
collaborations~\cite{Achasov:2002jv,Akhmetshin:2003rg,Ambrosino:2006hb}. 
In the case of charged pions, the  $e^+ e^-\to\gamma\pi^+\pi^-$
cross-section receives contributions  from both the initial-state radiation
(ISR) and from the FSR amplitudes. Interference effects
are sensitive to both the modulus and the phase of the FSR amplitude
but no model independent extraction has been attempted yet. 
Instead, theoretical modelling of the FSR amplitudes may be used to
improve the precision of the determination of the ISR one and the
related measurement of the pion form
factor~\cite{Dubinsky:2004xv,Pancheri:2006cp}.     

In the present paper, we discuss the generalization of the application
of complex plane methods, which were used to describe pion pair
production by two real photons, to the case where one photon is virtual.
Unitarity of the $S$-matrix is the basis for a model
independent treatment of the final-state interaction and leads  to
the Fermi-Watson theorem (e.g.~\cite{fermiwatson}) in the elastic
scattering regime. 
From this point of view it would seem that $\gamma\gamma^*(\qdd)$ scattering
could be intrinsically different from $\gamma\gamma$ since, in the
former case, the Fermi-Watson theorem may not apply even at low
$\pi\pi$ energy\footnote{
Indeed, if $\qdd>4\mpid$, the virtual photon can decay into two pions 
and the unitarity relation involves two terms
instead of just one in the elastic scattering regime:
$\im\braque{\gamma\gamma^*\vert\pi\pi}=
\braque{\gamma\gamma^*\vert\pi\pi}\braque{\pi\pi\vert\pi\pi} +
\braque{\gamma^*\vert\pi\pi}\braque{\gamma\pi\pi\vert\pi\pi} $. This
was pointed out in ref.~\cite{Creutz:1969hv}.}
depending on the value of $q^2$. However, as shown by
Omn\`es~\cite{omnes58} a more powerful result obtains by combining
unitarity with analyticity properties of the $S$-matrix, which leads,
for partial waves, to integral equations of the Muskhelishvili
type~\cite{Muskhelishvili}. We restrict ourselves here to an energy range
where inelasticity may be neglected in $\pi\pi$ scattering
(i.e. $s_{\pi\pi}\lapprox 1$ $\hbox{GeV}^2$) in which case the
Muskhelishvili equation is solvable in closed form in terms of the
final-state rescattering phase-shifts and the left-hand cut part of
the amplitude. Application of the
Muskhelishvili-Omn\`es (MO) formalism to the case of real
photon-photon scattering was discussed a long time
ago~\cite{gourdinmartin,Carlson:1971pk,Babelon:1976ww}. This 
was reconsidered in refs.~\cite{Morgan:1991zx,Donoghue:1993kw} who
showed how to implement theoretical constraints from the chiral
symmetry of QCD by matching with the calculations of chiral perturbation
theory (ChPT) at NLO~\cite{Bijnens:1987dc,Donoghue:1988eea}. 
The pion electric and magnetic polarizabilities are specific
observables involved in these amplitudes at low energy.
The phenomenological inputs in the work~\cite{Morgan:1991zx,Donoghue:1993kw}
are restricted to the description of the left-hand cut in
terms of light vector as well as axial-vector resonances. They achieved
a fair description of the available experimental data. 
Recently, a set of hyperbolic dispersion relations was developed for
$\gamma\gamma\to\pi\pi$~\cite{Hoferichter:2011wk} which, in principle,
allows for a more fundamental description of the left-hand cut if
experimental data on $\gamma\pi\to\pi\pi$, $\gamma\pi\to\pi\pi\pi$ were
available.  

The widths of the resonances can safely be ignored in the computation
of the left-hand cut for $\gamma\gamma$ scattering, but this is no
longer the case if one (or two) photons are virtual and $q^2 >0$
since, for large enough values, the resonances may be produced on
shell. The main issue, however, which we discuss in some detail is
whether the MO method is applicable at all in this regime. This is
because the left and right-hand cuts of the amplitude
$\gamma\gamma^*(\qdd)\to\pi\pi$ when $\qdd >4\mpid$ are no longer well
separated. The left-hand cut extends into the complex plane and intersects
and overlaps with the unitarity cut. We will show that this problem is
resolved using a proper description for the propagator of a finite
width resonance as well as a consistent application of limiting
$i\epsilon$ prescriptions for the energy variables.

The couplings of off-shell photons to hadrons involve
form factors. For this reason, we will consider here only the
contributions to the left hand cut generated by the vector mesons
$\rho$, $\omega$ (in addition to the pion pole contribution). In this
manner we have to deal with form factors for which some experimental
information is available. As usual with dispersive representations, it
is necessary to introduce polynomial subtraction parameters and we
assume that the contributions from heavier resonances can be
represented in this way in a restricted energy region. In the present
case, these parameters are actually functions of the photon
virtuality, $\qdd$. We discuss the constraints arising from the soft
photon as well as the soft pion limits. 
Our main result is an expression for the helicity amplitudes
$\gamma\gamma^*(\qdd)\to\pi\pi$ (or
$\gamma^*(\qdd)\to\gamma \pi\pi$) obeying these constraints
and in which the dependence on the $S$-wave $\pi\pi$ phase-shifts,
is displayed explicitly, as well as the dependence on the
$\pi\pi$, $\omega\pi$ and $\rho\pi$ electromagnetic form factors. This
expression is valid for negative as well as vanishing or positive
values of $\qdd$ and involves two unknown functions, $b^I(\qdd)$. 
Because of the restriction to the elastic $\pi\pi$ rescattering
region, the range of applicability is $\vert\qdd\vert < 1$ $\hbox{GeV}^2$. 
Comparing with the experimental
results of refs.~\cite{Achasov:2002jv,Akhmetshin:2003rg}  
a determination of $b^I(\qdd)$ in terms of the
pion polarizabilites and simply two real parameters is obtained. 

The plan of the paper is as follows. After introducing some notation
and useful kinematic formulae (sec.~\sect{basicformules}), we address
the problem of generalizing the left-hand cut structure arising from
resonance exchange contributions (sec.~\sect{lefthandcut}). In
sec.~\sect{omnesrep} we establish the dispersive MO representation for
the $J=0$ $\gamma\gamma^*$ partial-wave (which is the most
relevant in the  energy region considered). Then, (sec.~\sect{compexp})
we compare the resulting amplitudes with the available experimental
data, which determines the amplitudes completely. A few applications
of these amplitudes are presented, finally, in
sec.~\sect{applications}: we calculate, in particular, the generalized
pion polarizabilites, as introduced in
refs.~\cite{Unkmeir:1999md,L'vov:2001fz}. Concerning the $g-2$ of the
muon, we provide an evaluation of the contributions from the hadronic
vacuum polarization (HVP) associated with the state
$\gamma\pi^+\pi^-$ (which goes beyond the usual scalar QED approximation)
and from  $\gamma\pi^0\pi^0$.

\section{Basic formulae and notation}\lblsec{basicformules}
Let us consider the final-state radiation annihilation amplitude, 
$e^+(k_2)\,e^-(k_1)\to \gamma^*(q_2)\to $ 
$\gamma(q_1)\,\pi(p_1)\,\pi(p_2)$.  It
can be expressed as follows, 
\be\lbl{eplusemoins}
{\cal T}= e^3 \bar{v}(k_2)\gamma_\lambda u(k_1) \,
\left( g^{\lambda\nu}+(\xi-1){q_2^\lambda q_2^\nu\over \qdd} \right)
{1\over \qdd }\, W_{\mu\nu}(q_i,p_i) \epsilon_1^{*\mu}(q_1,\lambda_1)\ ,
\en
where $\epsilon_1$  is the polarization vector of  the photon and we
have denoted $q_2^2=(k_1+k_2)^2\equiv q^2$. 
An arbitrary  gauge parameter  $\xi$ was introduced in the
propagator of the off-shell photon. The tensor $W_{\mu\nu}$ is defined
from the following matrix element involving the  T-product of two
electromagnetic currents  
\be
e^2 W_{\mu\nu}(q_i,p_i)=i\int d^4x e^{-iq_1 x}
\braque{\pi(p_1)\pi(p_2)\vert T( j_\mu(x) j_\nu(0))\vert 0}\ .
\en 
Current conservation, i.e. $\partial_\mu j^\mu (x)=0$, leads
to the two Ward identities      
\be\lbl{wardid} 
q_1^\mu W_{\mu\nu}=0,\quad q_2^\nu W_{\mu\nu}=0\ .
\en
\subsection{Tensorial decomposition}
The  Ward  identities~\rf{wardid}   imply  that  $W_{\mu\nu}$  can  be
expanded, a priori, in terms of five independent
tensors~\cite{Bardeen:1969aw}  
$T_{i\mu\nu}$ made from the three independent momenta $q_1$, $q_2$,
$\Delta\equiv p_1-p_2$ 
and which satisfy the conditions~\rf{wardid}. Two of
these tensors give a vanishing contribution when contracted with the
photon polarization vector $\epsilon_1$ using $q_1^2=0$ and can be
ignored, such that one can write
\be
 W_{\mu\nu}(q_i,p_i)= A(s,t,u,\qdd) T_{1\mu\nu}+ B(s,t,u,\qdd) T_{2\mu\nu}+
 C(s,t,u,\qdd) T_{3\mu\nu} 
\en
where $s$, $t$, $u$ are the Mandelstam variables  
\be
s=(p_1+p_2)^2,\quad t=(p_1+q_1)^2,\quad u=(p_2+q_1)^2
\en
satisfying
\be
s+t+u=2\mpid+q^2\ .
\en
The three relevant tensors can be taken as\footnote{The first two
tensors are the same as used in ChPT calculations of
$\gamma\gamma\to\pi\pi$~\cite{bellucci,burgi}. The correspondence with
the tensors used in ref.~\cite{Dubinsky:2004xv} is as follows:
$T_1^{\mu\nu}=-\tau_1^{\nu\mu}$,
$T_2^{\mu\nu}=-4\tau_2^{\nu\mu}$, 
$T_3^{\mu\nu}= 2\tau_3^{\nu\mu}$.}
\bea
&& T_{1\mu\nu}= q_1\cdot q_2\, g_{\mu\nu} 
            -q_{1\nu} q_{2\mu}
\nonumber \\
&& T_{2\mu\nu}= 4\Delta_\mu     (q_1\cdot q_2\,
\Delta_\nu-q_2\cdot\Delta\, q_{1\nu}) 
            -4q_1\cdot\Delta (q_{2\mu}\Delta_\nu
            -q_2\cdot\Delta\, g_{\mu\nu}) 
\nonumber \\
&& T_{3\mu\nu}= 2\Delta_\mu    (q_1\cdot q_2\,  q_{2\nu}-q^2\, q_{1\nu})
            -2q_1\cdot\Delta(q_{2\mu} q_{2\nu} - q^2\, g_{\mu\nu})\ .
\ena
We note here that because of Bose symmetry of the $\pi\pi$ system with
$I=0,2$, the amplitudes must be invariant under interchange of the two
pion momenta $p_1$, $p_2$. The two tensors $T_{1\mu\nu}$ and
$T_{2\mu\nu}$ are even when $p_1\leftrightarrow p_2$ while the third
tensor $T_{3\mu\nu}$ is odd. This implies that the two functions $A$
and $B$ must be even under interchange of the two Mandelstam variables
$t$, $u$ while the function $C$ must be odd.

%%%%%%%%%%%%%%%%%%%pipi CMS figure
\begin{figure}[ht]
\bc
\includegraphics[width=0.6\linewidth]{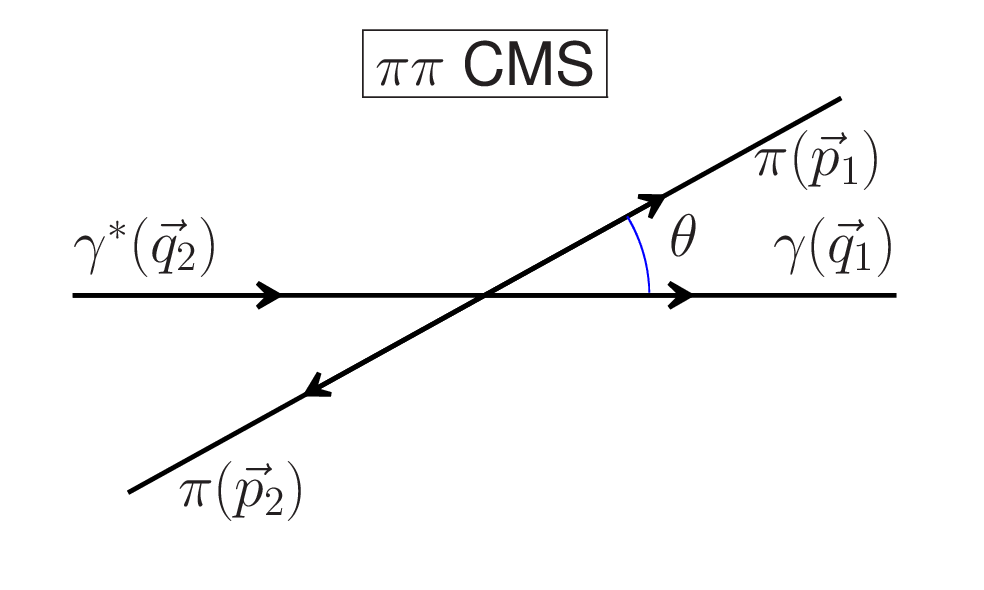}
\caption{\sl $\pi\pi$ center-of-mass system.}
\label{fig:pipicms}
\ec
\end{figure}
%%%%%%%%%%%%%%%%%%%pipi CMS figure
\subsection{Expressions in terms of helicity amplitudes}
Let us introduce a polarization vector $\epsilon_2(q_2,\lambda_2)$
associated with the virtual photon. We may then define a helicity  amplitude
$H_{\lambda_1,\lambda_2}$ by contracting the tensor
$W_{\mu\nu}(q_i,p_i)$ with the two polarization vectors,   
\be\lbl{Hlamb1lamb2}
e^{i(\lambda_2-\lambda_1)\phi}\,H_{\lambda_1,\lambda_2}(s,\qdd,\theta)\equiv
-W_{\mu\nu}(q_i,p_i)  
\epsilon_1^{*\mu}(q_1,\lambda_1)\epsilon_2^\nu(q_2,\lambda_2)\ .
\en
A minus sign is introduced here such that the limit $q^2=0$
corresponds to the $\gamma\gamma\to\pi\pi$ amplitude as usually
defined. The angles $\phi$ and $\theta$ are defined in the $\pi\pi$
center-of-mass system (see fig.~\fig{pipicms}) and we have factored out
explicitly the dependence  on the azimuthal  angle $\phi$. 

The second Ward identity~\rf{wardid} shows that the $e^+e^-$
annihilation amplitude~\rf{eplusemoins} is independent of the gauge
parameter $\xi$. Taking $\xi=0$ and using the identity
\be
{q_2^{\alpha} q_2^{\nu}\over \qdd} - g^{\alpha\nu}
=\sum_{\lambda_2} \epsilon_2^{*\alpha}(q_2,\lambda_2)
\epsilon_2^{\nu}(q_2,\lambda_2) 
\en
we find that the  $e^+e^-$ amplitude~\rf{eplusemoins} can be expressed
very simply in terms of the helicity amplitude
$H_{\lambda_1\lambda_2}$ introduced above~\rf{Hlamb1lamb2}  
\be
{\cal T}=e^3 \bar{v}(k_2)\eslashstar(q_2,\lambda_2) u(k_1) \,{1\over\qdd} 
e^{i(\lambda_2-\lambda_1)\phi}\,H_{\lambda_1\lambda_2}(s,\qdd,\theta)\ .
\en
Helicity amplitudes are convenient for performing the partial-wave
expansion~\cite{Jacob:1959at}. In the present case, it reads,
\be\lbl{pwexpand}
 H^I_{\lambda_1\lambda_2}(s,\qdd,\theta)=
\sum_J (2J+1) h^I_{J,\lambda_1\lambda_2}(s,\qdd) d^J_{\lambda_1-\lambda_2,0}(\theta)
\en
where we have introduced a superscript $I$  which labels the  isospin
state of the $\pi\pi$  system. The  relation between the isospin
amplitudes and the amplitudes corresponding to a charged or neutral
pion pair is 
\be\lbl{Hisomatrix}
\left(\ba{r}
H^0_{\lambda\lambda'} \\
H^2_{\lambda\lambda'}\ea\right)=\mathbf{C}
\left(\ba{c}
\sqrt2 H^c_{\lambda\lambda'}\\
H^n_{\lambda\lambda'}
\ea\right)\, ,\ \mathbf{C}=\mathbf{C}^{-1}=
\left(\ba{rr}
-\sqrt{\frac{2}{3}} & -\sqrt{\frac{1}{3}}\\
-\sqrt{\frac{1}{3}} &  \sqrt{\frac{2}{3}}
\ea\right)
\en
Since a two photon state is even under
charge conjugation so must be the $\pi\pi$  system which implies that
the isospin must be $I=0,2$. Consequently, the sum in
eq.~\rf{pwexpand} runs over even  values of the angular momentum $J$.
Recalling the  action of the parity operator~\cite{Jacob:1959at}, 
\be
P\vert JM\lambda_1\lambda_2\rangle= (-)^J \vert JM-\lambda_1-\lambda_2\rangle
\en
and the property of the $d^J$ functions
\be
d^J_{\lambda_1-\lambda_2,0}=(-)^{\lambda_1-\lambda_2}d^J_{-\lambda_1+\lambda_2,0} 
\en
we find the following relations between the helicity amplitudes
\be
H^I_{++}= H^I_{--},\quad
H^I_{+-}= H^I_{-+},\quad
H^I_{+0}=-H^I_{-0}
\en
such that only three of them are independent. 
In the $\pi\pi$ CMS frame, the Mandelstam invariants read,
\be\lbl{t-u}
t,u=\mpid +{1\over2}(\qdd-s)(1\mp\betapi(s)\cos\theta) 
\en
with
\be\lbl{betapi}
\betapi(s)=\sqrt{1-{4\mpid\over s}}\ .
\en
Using also the explicit expressions for the momenta and the
polarization vectors in this frame, one can  derive  the relations
between  the helicity  amplitudes
$H_{\lambda_1\lambda_2}(s,\qdd,\theta)$ and the  coefficient
functions $A(s,t,\qdd)$, $B(s,t,\qdd)$ and $C(s,t,\qdd)$ and one finds 
\bea\lbl{helicrels}%sign from eq.(8) also included 
&& H_{++}= (\qdd-s)\Big[ 
                                         {1\over2}\, A(s,t,\qdd) 
-(s-4\mpid)\left(1-{\qdd\over s}\cos^2\theta\right)\, B(s,t,\qdd)\nonumber\\
&&\phantom{ H_{++}}     -\qdd\betapi(s)\cos\theta \,C(s,t,\qdd)\Big] 
\nonumber \\[0.2cm]
&& H_{+-}=  (\qdd-s)(s-4\mpid)\sin^2\theta\, B(s,t,\qdd)\\[0.2cm]
&& H_{+0}=  (\qdd-s)\sqrt{\qdd} \sqrt{s-4\mpid}\,
{\sin\theta\over\sqrt2}\left[ 2\betapi(s)\cos\theta\,B(s,t,\qdd)
  -C(s,t,\qdd) \right] \ .
\nonumber 
\ena
Let us make a remark on the behaviour of the amplitudes when the
energy of the $\pi\pi$ system is close to the threshold. Since $C$ is
an odd function of  $t-u$ we can denote
\be
C(s,t,\qdd)\equiv (t-u) \tilde C(s,t,\qdd)\ .
\en 
Using eq.~\rf{t-u} shows that $C(s,t,\qdd)$ should be proportional to
$\sqrt{s-4\mpid}$. It then follows from the expressions for the
helicity amplitudes~\rf{helicrels} that when $s\to 4\mpid$, the
amplitude $H_{++}$ remains finite while the other two helicity
amplitudes $H_{+-}$, $H_{+0}$ vanish as $O(s-4\mpid)$. This reflects
the fact that in these amplitudes the $\pi\pi$ pair must be in a
state of angular momentum $J\ge 2$. $H_{++}$ therefore dominates at
low $\pi\pi$ energies.   

%%%%%%%%%%%%%%%% e+e- CMS figure
\begin{figure}[ht]
\bc
\includegraphics[width=0.7\linewidth]{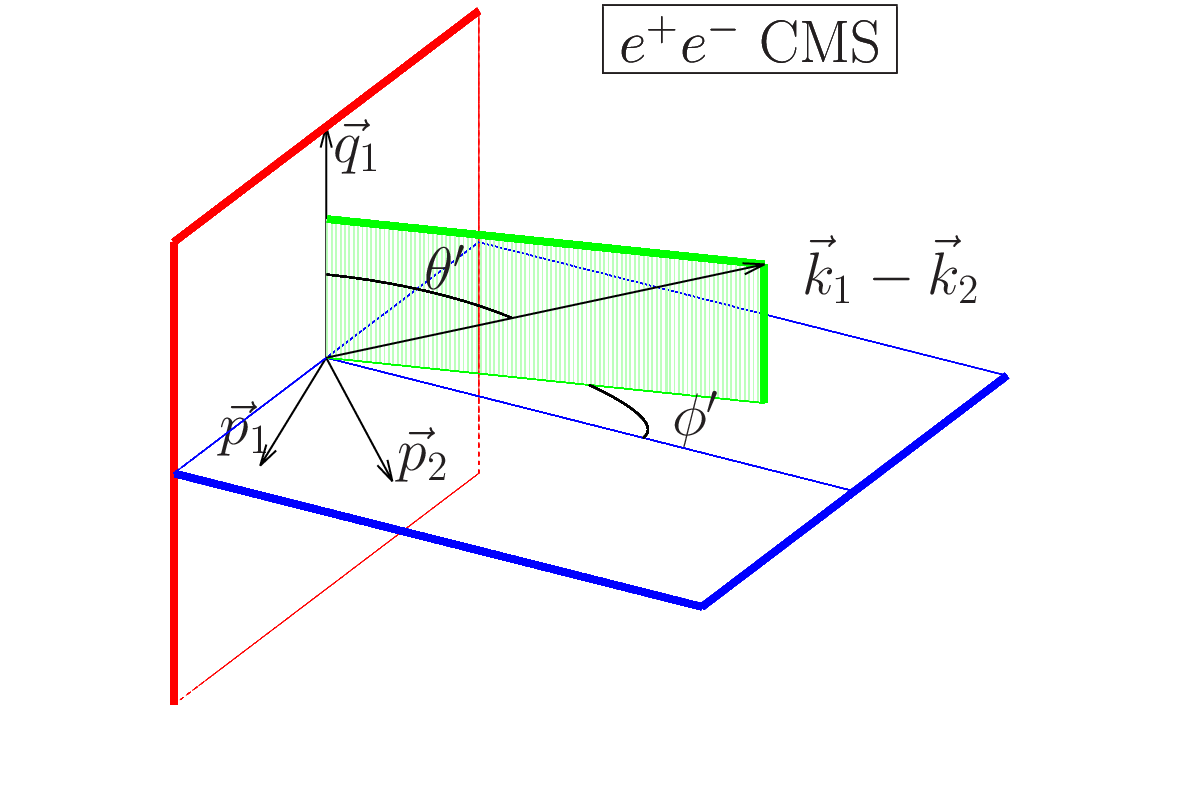}
\caption{$e^+ e^-$ center-of-mass system.}
\label{fig:eecms}
\ec
\end{figure}
%%%%%%%%%%%%%%%% e+e- CMS figure
\subsection{$e^+ e^-$ center-of-mass frame }
Let us now consider the CMS system of the $e^+ e^-$ pair i.e.
\be
\vec{k}_1+\vec{k}_2=\vec{q}_2=0
\en
The momenta of the photon and that of the two pions sum to zero 
$\vec{p}_1+\vec{p}_2+\vec{q}_1=0$ and therefore lie in a plane. This
plane is determined by two polar angles, which we call $\theta'$,
$\phi'$ with respect to the $e^+ e^-$ beam axis
$\vec{k}_1-\vec{k}_2$. This is illustrated in fig.~\fig{eecms}.
We can use $\vec{q}_1$ as $z$ axis and write $k_1-k_2$ in
terms of $\theta'$, $\phi'$
\be
k_1-k_2=\sqrt{\qdd-4\med}\left(\ba{c}
0  \\
\cos\phi'\sin\theta' \\
\sin\phi'\sin\theta' \\
\cos\theta'\\
\ea\right)\ .
\en
The momenta  and the polarization vectors in this new frame are
obtained from  those in the $\pi\pi$ CMS frame by acting with the Lorentz
transformation 
\be
L= \left(\ba{cccc}
\gamma &0&0& -\beta\gamma\\
0&1&0&0\\
0&0&1&0\\
-\beta\gamma &0&0& \gamma\\
\ea\right) \quad \hbox{with}\
\gamma={s+\qdd\over \sqrt{4 s \qdd}},\
\beta\gamma={\qdd-s\over \sqrt{4 s \qdd}}
\en
Useful scalar products involving $k_1-k_2$ are listed below
\bea\lbl{prodeps2}
&&(k_1-k_2)\cdot \epsilon_2(\lambda_2)=
\sqrt{\qdd-4\med} e^{-i\lambda_2 \phi'}\big(
\lambda_2{\sin\theta'\over\sqrt2} -\delta_{\lambda_2 0}\cos\theta' \big)
\nonumber\\
&& (k_1-k_2)\cdot q_2=0
\nonumber\\
&& (k_1-k_2)\cdot q_1=(s-\qdd)\sqrt{{\qdd-4\med\over4\qdd}}\cos\theta'
\nonumber\\
&& (k_1-k_2)\cdot\Delta= -\sqrt{(s-4\mpid)(\qdd-4\med)}\Big(
\sin\theta\sin\theta'\cos(\phi-\phi')
\nonumber\\
&&\phantom{(k_1-k_2)\cdot\Delta}
+ {s+\qdd\over\sqrt{4 s \qdd}} \cos\theta\cos\theta' \Big)\ .
\ena
The differential cross-section  for $e^+ e^-\to
\gamma(p_1,\lambda_1)\pi\pi$, assuming unpolarized $e^+ e^-$ beams, 
can be expressed as follows in terms of the helicity amplitudes
(using eqs.~\rf{prodeps2})
\bea
&& d^4\sigma_{\lambda_1}={e^6 \over 4\sqrt{\qdd(\qdd-4\med)}}
\left({1\over\qdd}\right)^2  \Bigg\{\qdd\,\left(
\abstwo{H_{\lambda_1+}}+\abstwo{H_{\lambda_1-}}+\abstwo{H_{\lambda_10}} \right)\\
&& -(\qdd-4\med)
\abstwo{{\sin\theta'\over\sqrt2}\left( 
 H_{\lambda_1+} e^{i(\phi-\phi')} -H_{\lambda_1-} e^{-i(\phi-\phi')}\right)
 -\cos\theta' H_{\lambda_10}}
\Bigg\}\, d\hbox{Lips}_3 \ .\nonumber
\ena
The three-body phase-space integration measure has the following
expression  in terms of the $\pi\pi$ energy $s$ and the angular
variables $\theta$,  $\theta'$, 
$\phi-\phi'$  
\be
d\hbox{Lips}_3 ={ (\qdd-s)\betapi(s)\over4 (4\pi)^4\,\qdd}
ds\,d\cos\theta\, d\cos\theta'\,d(\phi-\phi')\ . 
\en 
In practice, the distribution over  the Dalitz plot  is obtained after
integrating over the variables  $\theta'$, $\phi-\phi'$. This  partly
integrated cross-section, summed over the two photon helicities,  has
the following expression 
\be\lbl{d2sigma}
 {d^2\sigma\over ds\, d\cos\theta}= 
 {\alpha^3 (q^2+2m_e^2)(\qdd-s)\betapi(s)\over 
12\sqrt{\qdd(\qdd-4\med)}(q^2)^3 } 
\left( \abstwo{H_{++}}+\abstwo{H_{+-}}+\abstwo{H_{+0}}\right)
  \ .
\en
The cross-section $\sigma(\qdd)$ is obtained by integrating over
$\theta$ in the range $[0,\pi]$ for charged pions ($[0,\pi/2]$ for
neutral pions) and integrating over $s$ in the range $[4m_\pi^2,\qdd]$.

\section{A model for the (generalized) left-hand cut}\lblsec{lefthandcut}
In order to implement the MO method to the partial-wave amplitudes
$h^I_{\lambda\lambda'}(s,\qdd)$ we must consider the analytical
structure as a function of the $\pi\pi$ energy variable $s$, and input
a model for the left-cut part of the amplitude. This left-hand cut
originates from singularities (poles, cuts) of the unprojected amplitude as a
function of the Mandelstam variables $t$, $u$. A first contribution,
essentially model independent, arises from the charged pion pole. We will
then consider contributions associated with light vector
resonances. At first, we will ignore the widths of the resonances,
such that the contributions are also simple poles in the $t$, $u$
variables. Such an approximation is acceptable for $\gamma\gamma$ but
not for $\gamma\gamma^*$ with $\qdd > 4\mpid$. One problem which raises
is that the left-hand cut is no longer well separated from the
unitarity cut. As we will show below, a solution to this problem
consistent with expected general properties of the amplitude, is to
construct a resonance propagator which has a cut instead of a pole.

%%%%%%%%%%%%%%%%one-pion exchange figure
\begin{figure}[ht]
\bc
\includegraphics[width=12cm]{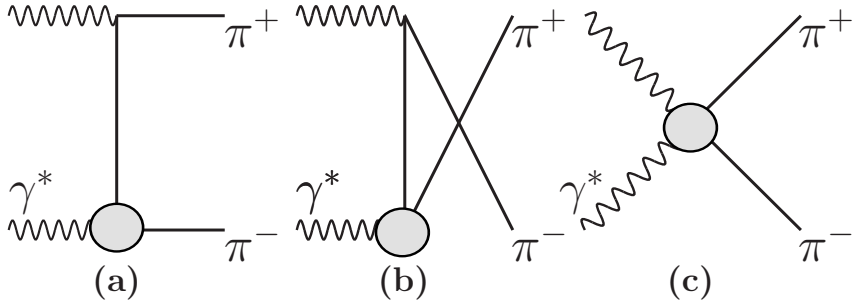}
\caption{\sl Born diagrams (one-pion exchange) contributions to the
  $\gamma\gamma^*\to \pi^+\pi^-$ amplitude.}
\label{fig:borndiags}
\ec
\end{figure}
%%%%%%%%%%%%%%%%one-pion exchange figure
\subsection{One-pion exchange (Born) amplitudes}
The diagrams for the charged pion pole amplitudes are shown in
fig.~\fig{borndiags}. The blobs indicate that the vertex must take
into account that the photon is off-shell. The 
matrix element of the electromagnetic   current between two pions
involves   the  pion form factor function $F^v_\pi(s)$,  
\be\lbl{ffactordef}
\braque{\pi^+(p)\vert j_\mu(0)\vert \pi^+(p')}=(p+p')_\mu F^v_\pi( (p-p')^2)
\en
We can use this matrix element to provide a {\sl definition} of the
pion pole contributions (diagrams (a), (b) in
fig.~\fig{borndiags}). The dependence of the vertex on the fact that
one pion is off the mass shell can absorbed into the non-pole contributions
(diagram (c)).
From diagrams (a), (b) one then obtains,
\be\lbl{bornab}
W^{Born,a+b}_{\mu\nu}= {F^v_\pi(\qdd)}\left[
{(\qdd-s)\,T_{1\mu\nu}-{1\over2} T_{2\mu\nu}\over (t-\mpid)(u-\mpid)}
  -2g^{\mu\nu}          \right]\ .
\en
expressed in  terms of the $T_{i\mu\nu}$ tensors,
which has exactly the same expression as in the case of on-shell
photons except that it is multiplied by the form factor $F^v_\pi(\qdd)$. 
Gauge invariance dictates that the contribution from the diagram (c)
must cancel the last term in eq.~\rf{bornab}.  Of course, there can be
additional, gauge invariant contributions from this class of
diagrams. Some of them, which can be associated with $\rho$, $\omega$
vector resonance exchanges will be considered below. 
In a dispersive approach, further contributions are absorbed into
subtraction functions. The Born terms, finally, can be defined as, 
\be\lbl{Wborn}
W^{Born}_{\mu\nu}= A^{Born}(s,t,\qdd)\,T_{1\mu\nu}+B^{Born}(s,t,\qdd)\,T_{2\mu\nu}
\en
with
\be\lbl{ABborn}
A^{Born}(s,t,\qdd)= {F^v_\pi(\qdd)(\qdd-s)\over (t-\mpid)(u-\mpid)},\
B^{Born}(s,t,\qdd)= {-F^v_\pi(\qdd)\over 2(t-\mpid)(u-\mpid)}\ .
\en
Next, using the relations~\rf{helicrels}, the three helicity amplitudes
corresponding to the Born diagrams can be deduced. Using eq.~\rf{pwexpand},
we can compute the partial-waves, the $J=0$ partial-wave amplitude reads,
\be
\lbl{pwborn0}
 h^{Born}_{0,++}(s,\qdd)={F^v_\pi(\qdd)\over s-\qdd}\,
\left[
4\mpid\, {L_\pi(s)}  -2\qdd \right],\quad
L_\pi(s)= {1\over \betapi(s)}
\log{ 1 + \betapi(s)\over1 -\betapi(s)}\ .  
\en
The corresponding isospin $I=0,\ 2$ amplitudes,using~\rf{Hisomatrix}
are given by 
\be\lbl{isoborn}
 h^{I,Born}_{0,++}(s,\qdd)=-\sqrt{4-I\over 3}
 h^{Born}_{0,++}(s,\qdd)\ .
\en

Let us examine the singularities of $h^{Born}_{0,++}(s,\qdd)$ in the
complex plane of the variable $s$. The function $L_\pi(s)$ has a
singularity on the negative real axis $s\in [-\infty,0]$, which is
the expected left-hand cut. In addition, if $\qdd\ne0$, there is a pole
singularity when $s=\qdd$. This value of $s$ corresponds to the
kinematical situation where the real photon becomes soft
i.e. $q_1\to0$ as one can see from the relation
\be
\qdd-s=2q_1\cdot q_2\ .
\en
When $\qdd > 4\mpid$, this singularity overlaps with the unitarity
cut. However, as $\qdd$ is an energy variable (it is the invariant
energy of the $e^+ e^-$ pair) the amplitudes must be defined with the
$i\epsilon$ limiting prescription i.e. $\qdd=\lim
\qdd+i\epsilon$. This prescription shifts the pole singularity away
from the unitarity cut. 
%%%%%%%%%%%%%%%%%%%%%vector meson exchange figure
\begin{figure}[ht]
\bc
\includegraphics[width=12cm]{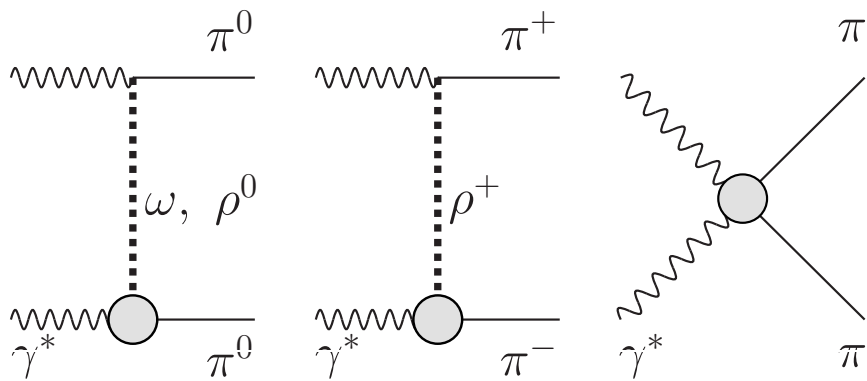}
\caption{Vector-meson exchange diagrams and possible associated
  contact term.}
\lblfig{VPgraphs}
\ec
\end{figure}
%%%%%%%%%%%%%%%%%%%%%vector meson exchange figure
\subsection{Vector-meson exchange amplitudes in the zero-width limit}\lblsec{vectorex}
The  diagrams corresponding  to charged and neutral vector  meson
exchanges  are shown in fig.~\fig{VPgraphs}. At first, let us ignore
the widths of the resonances. We start from the following Lagrangian which
describes the coupling of a real photon to a vector meson and a pion,
\be\lbl{LVPgamma}
{\cal L}_{VP\gamma}= eC_V \epsilon^{\mu\nu\alpha\beta} F_{\mu\nu} 
\partial_\alpha P V_\beta \ .
\en
(where $P$ is either $\pi^0$ or $\pi^\pm$ depending on the charge of
$V$). Thanks to the derivative couplings the amplitude computed
from~\rf{LVPgamma} should automatically vanish in the soft photon
limit as well as in the soft pion limit.
The coupling constants $C_V$ are related directly to the decay widths
of the vector mesons into $P \gamma$, 
\be
\Gamma_{V\to P\gamma}=\alpha\,{\tilde C}_V {(\mvd-m_P^2)^3\over 3
  m_V^3}
\en
with
\be
\tilde{C}_V= {1\over2}C_V^2\ .
\en
The following numerical values for the couplings ${\tilde C}_V$ can be
deduced from the PDG~\cite{Beringer:1900zz}
\be\lbl{CtildeV}
\ba{ll}
\Gamma(\omega\to \pi^0\gamma)=703\pm25\ \hbox{KeV}\quad& 
{\tilde C}_\omega= 0.66\pm0.023\ (\hbox{GeV}^{-2}) \\
\Gamma(\rho^0\to \pi^0\gamma)=89\pm12 \  \hbox{KeV}\quad&
{\tilde C}_{\rho^0}=0.09\pm 0.01 \ (\hbox{GeV}^{-2}) \\
\Gamma(\rho^+\to \pi^+\gamma)=68\pm7 \  \hbox{KeV}\quad&
{\tilde C}_{\rho^+}=0.07\pm 0.007 \  \hbox{GeV}^{-2}\ . \\ 
\ea
\en
%modif here:
When  the photon  is off-shell, the vertex  is modified  by  a 
form factor $F_{V\pi}$ which can be defined, in the zero width
limit, from the matrix element 
\be\lbl{Vpiffdef}
\braque{V(\lambda,p_V)\vert j_\mu(0)\vert \pi(p_\pi)}=
2eC_V\,F_{V\pi}(\qdd)\, \epsilon_{\mu\alpha\beta\gamma} p_V^\alpha p_\pi^\beta
\varepsilon_V^{*\gamma}(\lambda) 
\en
and is normalized such  that $F_{V\pi}(0)=1$.  Computing
the amplitude from the first two diagrams of fig.~\fig{VPgraphs} with
this vertex, one finds
%sign adapted to final-state gamma
\bea\lbl{Wvector}
&& A^V(s,t,\qdd)=\tilde{C}_V F_{V\pi}(\qdd)\left[
 {s-4m_\pi^2-4t+\qdd\over t-\mvd}
+{s-4m_\pi^2-4u+\qdd\over u-\mvd}\right]    \nonumber\\
&& B^V(s,t,\qdd)=\tilde{C}_V F_{V\pi}(\qdd)\left[
{1\over2(t-\mvd)} +{1\over2(u-\mvd)}\right] \nonumber\\
&& C^V(s,t,\qdd)=\tilde{C}_V F_{V\pi}(\qdd)\left[
{1\over t-\mvd} -{1\over u-\mvd}\right]\ .
\ena
We have used the fact that upon interchanging $p_1$, $p_2$ the tensors
$T_1^{\mu\nu}$,   $T_2^{\mu\nu}$ are left invariant while
$T_3^{\mu\nu}\to -T_3^{\mu\nu}$ and $t,\ u$ are interchanged. We note
that the expression of the function $A^V$ involves a linear off-shell
dependence in $t$, $u$ in the numerators of the pole. It is possible,
in principle, to replace $t$, $u$ by $\mvd$ by adding a contribution
from the non-pole diagram in fig.~\fig{VPgraphs}. Doing this, however,
would spoil the correct soft pion limit. We can now compute the
corresponding vector-exchange helicity amplitudes and project on the
partial waves. We obtain for $J=0$,  
\bea\lbl{pwvec0}
&&{h}^V_{0,++}(s,\qdd,\mvd)= \tilde{C}_VF_{V\pi}(\qdd)\bigg\{
{L_V(s,\qdd,\mvd)\over \betapi(s)}
\bigg[ -4m_V^2 +4\qdd\left({\mvd-\mpid\over s-\qdd}\right)^2\bigg] 
\nonumber\\
&&+2\qdd \left(1- {2(\mvd-\mpid)\over s-\qdd}\right)
 +4(s-\qdd) \bigg\}
\ena
with
\be\lbl{Lv}
L_V(s,\qdd,\mvd)= \log( {\mvd -t^+(s,\qdd))-\log(\mvd-t^-(s,\qdd)})
\en
and
\be
t^\pm(s,\qdd)=\mpid +{1\over2}(\qdd-s)(1\pm \betapi(s))\ .
\en

\subsection{Complex singularity structure of the vector meson exchange
  amplitudes} 
Let us now consider the singularities of the partial-wave
amplitude~\rf{pwvec0}. Contrary to the case of  pion exchange,
the vector amplitude has no pole in the soft photon limit $s=\qdd$. In
fact it is easy to verify that the partial-wave amplitude~\rf{pwvec0}
vanishes at this point.  The cuts now are contained in the function
$L_V(s,\qdd,\mvd)/\betapi(s)$. Concerning the branch points, in
addition to the points $s=0$, $s=\infty$ there are two finite branch points     
\be\lbl{splusmoin1}
s_\pm(\qdd,\mvd)= \qdd -{\mvd-\mpid\over2\mvd}\left(\qdd+\mvd-\mpid\mp
\lambda^{1\over2}(\qdd,\mvd,\mpid) \right)
\en 
with
\be
\lambda(\qdd,\mvd,\mpid)=(\qdd-\mmd)(\qdd-\mpd),\quad m_\pm = m_V\pm
m_\pi\ .
\en
An alternative useful expression for these branch points can be derived,
\be\lbl{splusmoin2}
s_\pm = {4m_\pi^2\over (A\mp B)^2},
\en
with
\be
A= {m_+\over\sqrt{\qdd}}\sqrt{1-{m_-^2\over\qdd}},\quad
B= {m_-\over\sqrt{\qdd}}\sqrt{1-{m_+^2\over\qdd}},\quad
\en
Depending on the value of $\qdd$ one has to consider three cases
\begin{enumerate}
\item $\qdd < m_-^2$: In this case,  both $A$ and $B$ are imaginary,
  $s_\pm$  are then real and lie on the negative axis. The real cut,
  in this situation, is entirely situated on the negative real axis
  and consists of the two pieces $[-\infty,s_-]$, $[s_+,0]$. 
  
\item $m_-^2 \le \qdd \le m_+^2$: in this case the branch points are
  complex, the real cut consists of the entire negative axis $[-\infty,0]$ 

\item $\qdd > m_+^2$:  In this case, the branch points are again real
 and since both $m_+\over\sqrt{\qdd}$ and
 $m_-\over\sqrt{\qdd}$ are smaller than one, we can set 
\be
{m_+\over\sqrt{\qdd}}\equiv \sin{a},\quad
{m_-\over\sqrt{\qdd}}\equiv \sin{b}
\en
such that one can express $s_\pm$ as
\be
s_\pm = {4\mpid\over \sin(a\mp b)^2}\ .
\en
This expression shows that  both branch points are real and
larger than $4\mpid$.
\end{enumerate}
In addition to the cuts on the real axis, the meson
exchange amplitudes  have a complex cut corresponding to complex $s$
solutions of the equations $\im((\mvd-t^+)/(\mvd-t^-))=0$ , $\re
((\mvd-t^+)/(\mvd-t^-)) \le 0$.  These complex cuts are illustrated in
fig.~\fig{Lvcuts}. In the case when $\qdd > (m_V+m_\pi)^2$ the figure
shows that the complex cut intersects the unitarity cut.
%%%%%%%%%%%%%%complex cut for zero width figure
\begin{figure}[ht]
\bc
\includegraphics[width=0.5\linewidth]{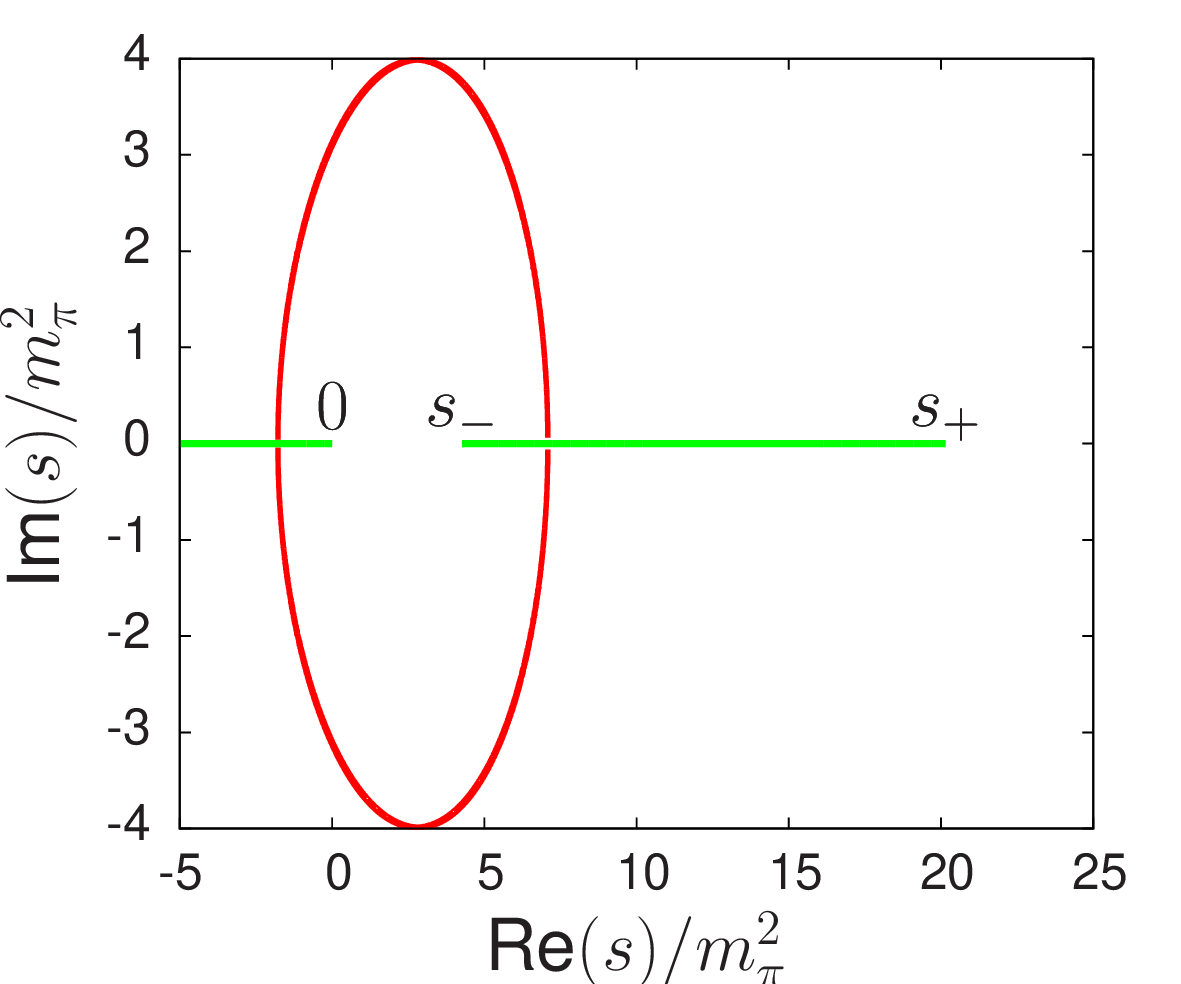}
\caption{Real and complex cuts of the function $L_V(s,\qdd,\mvd)$, with
  $m_V=0.77$ GeV and $\qdd=1\ \hbox{GeV}^2$.}
\lblfig{Lvcuts}
\ec
\end{figure}
%%%%%%%%%%%%%%complex cut for zero width figure
\subsection{Finite width resonance exchange amplitudes with correct
  analyticity properties}\lblsec{bwtilde}
Applicability of the MO method to the amplitudes of interest relies on
the ability to separate the amplitude (via a Cauchy representation)
into a piece having only a left-hand cut and a piece having only a
right hand one. This, a priori, is not the case if the left-hand cut
is of the form illustrated in fig.~\fig{Lvcuts}. 
So far, however, we have  ignored the width of the resonances i.e. we
have taken the propagator to be a simple pole in the variables $t$,
$u$. A naive way of trying to solve the problem is to use a complex
resonance mass, i.e. replace $m_V$ by $m_V-i\Gamma_V/2$, but this
ansatz is not quite correct since the resonance pole lies on the first
sheet. A related issue is that the propagator would be complex independently
of the value of $t$, $u$. The correct analyticity properties expected
from a resonance propagator are that the resonance poles should be
located on the second Riemann sheet and that the propagator be
analytic as a function of its variable  except for a right-hand cut.
Consider specifically the $\rho$ meson,
whose width is dominated by two-particle decay. Propagators which are
currently used do not exactly satisfy these properties. For instance,
the  Breit-Wigner propagator with a momentum dependent width 
\be
BW_V(t)={1\over m_V^2-t -i\gamma_V \betapi(t)(t-4\mpid)}\ ,\quad 
\gamma_V={m_V \Gamma_V\over \betapi(\mvd)(\mvd-4\mpid)}
\en
has a left-hand cut. The propagator proposed by Gounaris and
Sakurai~\cite{GS}  has no left-hand cut but  has an unphysical
pole singularity on the real axis\footnote{In the case of the 
$\rho$ meson parameters, this is  a formal rather than a practical
problem, as the singularity is located at a rather large negative
value $t\simeq -9.4\,10^5$ GeV$^2$.}.  
On rather general grounds, one expects  that a propagator should
satisfy a K\"allen-Lehmann dispersive representation~\cite{KallenLehmann}, 
\be\lbl{spectralprop}
\widetilde{BW}_V(t)= {1\over\pi}\int_{4\mpid}^\infty dt'\, 
{\sigma(t',m_V,\Gamma_V)\over (t'-t)}\ 
\en 
which automatically ensures the absence of singularity in the complex plane
except for a right-hand cut.      
This propagator is well defined and real when $t$ is real and smaller
than $4\mpid$, unlike $BW_V(t)$. For the spectral function
$\sigma(t',m_V,\Gamma_V)$ we can use, for instance, the imaginary part
of the BW propagator 
\be\lbl{kstilde}
\sigma(t',m_V,\Gamma_V)=\im[BW_V(t')]\ 
\en
which is definite positive 
(this ansatz has been considered before, e.g.~\cite{Lomon:2012pn}). 
In this case, $\widetilde{BW}_V(t)$ and $BW_V(t)$ have the same imaginary
parts when $t\ge 4\mpid$ but they have (slightly) different real parts
(see appendix~\sect{Vpropagator}). 

%%%%%%%%%%%%%%%%cut for finite width
\begin{figure}[ht]
\bc
\includegraphics[width=6cm]{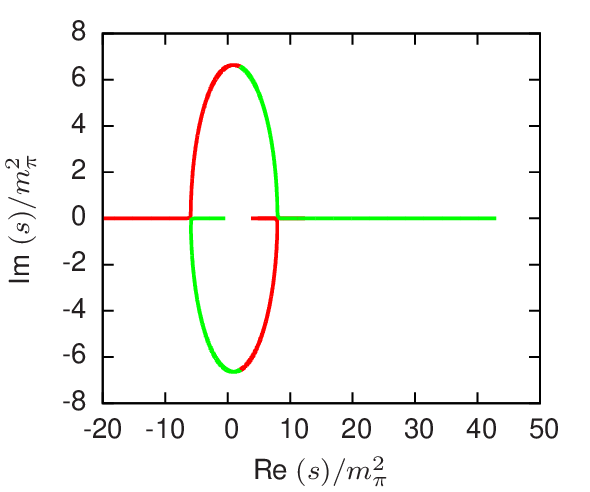}
\includegraphics[width=6cm]{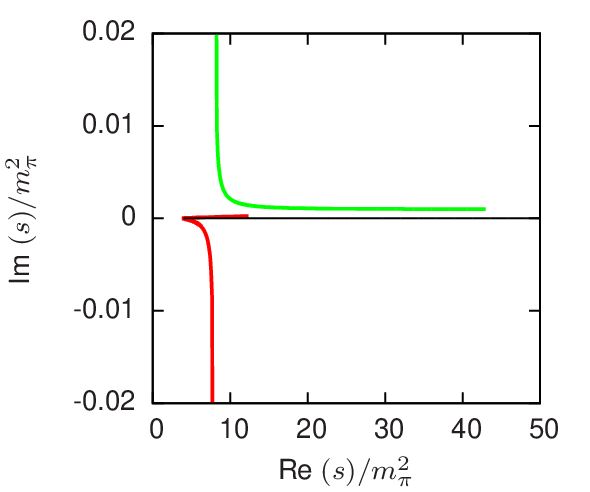}
\caption{Real and complex cuts of a  resonance exchange amplitude
  using a spectral representation of the propagator
  (see~\rf{spectralprop}, \rf{Ltilde}. The figure on the right
illustrates how this cut avoids the unitarity cut.}
\lblfig{Ltildecut}
\ec
\end{figure}
%%%%%%%%%%%%%%%%cut for finite width
%modif here
Let us then assume a phenomenological treatment of finite width
effects in the vector meson exchange amplitudes restricted to a simple
replacement of the propagators\footnote{This ansatz preserves the
  correct soft photon and soft pion limits.}
\be\lbl{widthreplace}
{1\over m_V^2-z}\longrightarrow \widetilde{BW}_V(z),\ z=t,\,u
\en
in eq.~\rf{Wvector}, where $\widetilde{BW}_V(z)$ has the dispersive
form given in eq.~\rf{spectralprop}. 
The corresponding partial-wave amplitudes can then be expressed in the form of
spectral representations and the cuts of the partial-wave amplitudes
are contained in the function $\tilde{L}_V$ which is given by
\bea\lbl{Ltilde}
&& \tilde{L}_V(s,\qdd,m_V,\Gamma_V)=\nonumber\\
&&\qquad \int_{4\mpid}^\infty dt'
      \sigma(t',m_V,\Gamma_V) \left[
	\log(t'-t^+(s,\qdd))-\log(t'-t^-(s,\qdd)) \right]\ .
\ena
The equations for the cuts, in the case of a
representation~\rf{Ltilde} are given in parametric form by the locus
of the singularities of the logarithms i.e. $s^\pm(\qdd,t')$ with
$4\mpid\le  t'\le \infty$. One obtains the same result as in a more general
derivation relying on the Mandelstam double spectral representation
of the amplitude~\cite{kennedyspearman}. It would seem that, again,
the cut  intersects and overlaps  the unitarity cut. However, one must
remember that $q^2$  must be considered as a limiting value of
$q^2+i\epsilon$. Figure~\fig{Ltildecut} shows the global shape of the
cut and a more detailed view of the vicinity of the unitarity cut
using the $\qdd+i\epsilon$ prescription. The figure shows that the
cut has two branches: the upper branch of the cut lies strictly above
the unitarity cut while the lower branch crosses the real axis close
to $4\mpi^2$. A simple calculation shows that this crossing occurs at
the point    
\be
s_c= 4\mpid\left(1-{\epsilon^2\over(\qdd-4\mpid)^2} \right)
\en
(corresponding to the parameter value
$t'=\qdd/2-\mpid$) which is located strictly below $4\mpid$. In
conclusion, the cuts of the resonance amplitude are now definitely
separated from the unitarity cut and this implies that the usual
MO method is applicable. 

%modifs: rewriting of beginning of sec. 4
%note added after eq. (58)
\section{Dispersive Omn\`es representations }\lblsec{omnesrep}
The discussion above justifies that the usual Omn\`es dispersive
representation~\cite{omnes58} applies to the partial-wave
$\gamma\gamma^*$ amplitudes $h^I_{J,\lambda\lambda'}(s,\qdd)$
in terms of the (generalized) left-cut part and the Omn\`es
function $\Omega_J^I(s)$. The form of this representation, in which
the dependence on the two variables $s$, $\qdd$ is displayed
explicitly, is given below~\rf{OMformule}. As a check of its
correctness, i.e. of the absence of an anomalous threshold and that
the real and imaginary parts are correctly computed, we consider in
appendix~\sect{Disptriangles} a toy model of rescattering, which leads
to simple triangle diagrams which can be computed in two different
manners. 

It is convenient to display explicitly the pole at $s=\qdd$ of the
Born term as well as  the form factor $F^v_\pi(\qdd)$ 
\be
h^{I,Born}_{0,++}(s,\qdd)\equiv {F^v_\pi(\qdd)\over s-\qdd}\,
\,\bar{h}^{I,\pi}_{0,++}(s,\qdd).
\en
Similarly, the form factor $F_{V\pi}(\qdd)$ may be displayed in the case of the
vector meson exchange amplitudes, in the zero width limit at first,
\be\lbl{hbarivdef}
h^{I,V}_{0,++}(s,\qdd,\mvd)\equiv
F_{V\pi}(\qdd)\,\bar{h}^{I,V}_{0,++}(s,\qdd,\mvd)\ .
\en 
In the finite width case, using eq.~\rf{widthreplace}, the following spectral
representation holds\footnote{ In practice, this
representation was used  only for the $\rho$-meson, the spectral
integration being performed numerically. Alternatively,  the
computation of the partial-wave amplitudes can be done starting
from the finite width helicity amplitudes,
eqs.~\rf{Wvector},~\rf{widthreplace},  
and performing the angular integration numerically, which
provides a check on the calculation. In the
case of the $\omega$ meson, the finite width was implemented more
naively by using a complex mass, i.e. setting $\mvd\equiv
(m_\omega-i\Gamma_\omega/2)^2$ in $\bar{h}^{I,V}_{0,++}(s,\qdd,\mvd)$. } 
\be\lbl{spectralpropamplitbar}
\tilde{h}^{I,V}_{0++}(s,\qdd,m_V,\Gamma_V)={1\over\pi}\int_{4\mpid}^\infty
dt'\sigma(t',m_V,\Gamma_V) \bar{h}_{0++}^{I,V}(s,\qdd,t')\ .
\en 
Let us then  introduce the following two integrals, 
\bea\lbl{Jintegrals} 
&& J^{I,\pi}(s,\qdd)={1\over\pi}\int_{4\mpid}^\infty {ds'\over
  (s')^2(s'-s)}{\sin\delta_0^I(s')\over\vert\Omega_0^I(s')\vert}\,
  \bar{h}^{I,\pi}_{0,++}(s',\qdd)\nonumber\\
&& J^{I,V}(s,\qdd)={1\over\pi}\int_{4\mpid}^\infty {ds'\over
  (s')^2(s'-s)}{\sin\delta_0^I(s')\over\vert\Omega_0^I(s')\vert}\,
\tilde{h}^{I,V}_{0,++}(s',\qdd,m_V,\Gamma_V)\ ,
\ena
where $\Omega_0^I(s')$ is given in eq.~\rf{omnesrecall}. It is worth
noting here that  the $s'$ integrations in eq.~\rf{Jintegrals} are
well defined when using the $s+i\epsilon$ and $\qdd+i\epsilon$
prescriptions, since the singularities from $1/(s'-s)$ and from
$\tilde{h}^{I,V}_{0,++}(s',\qdd,m_V,\Gamma_V)$ are then moved away from the
real axis as has been discussed above in sec.~\sect{bwtilde}. 

Writing dispersion relations with  two subtractions at $s=0$, the
representation for the $H_{++}$ helicity amplitude, taking into
account rescattering in the $S$-wave, can then be written as 
\bea\lbl{OMformule}
&& H^I_{++}(s,\qdd,\theta)=
F^v_\pi(\qdd)\bar{H}^{I,Born}_{++}(s,\qdd,\theta)
+\sum_{V=\rho,\omega} F_{V\pi}(\qdd)\bar{H}^{I,V}_{++}(s,\qdd,\theta)
\nonumber\\
&&\quad +\Omega_0^I(s)\Bigg[ a^I(\qdd) +s\,b^I(\qdd) +s^2\,F^v_\pi(\qdd)\,
{J^{I,\pi}(s,\qdd)- J^{I,\pi}(\qdd,\qdd)\over s-\qdd}
\nonumber\\
&&\quad +s^2\,\sum_{V}F_{V\pi}(\qdd)\,J^{I,V}(s,\qdd)\Bigg]\ .
\ena
We have indicated explicitly that the two
subtraction constants $a^{I}$ and $b^{I}$ should depend on $\qdd$. These
functions are expected to be analytic as a function of $\qdd$ with a cut
on the real axis, $\qdd > 4\mpid$. When $\qdd$ is real and smaller
than $4\mpid$, $a^{I}$ and $b^{I}$ are real as well as the form
factors, which ensures that the partial wave amplitude from the
representation~\rf{OMformule} satisfies Watson's theorem.
Let us now consider the constraints arising from the soft-photon and
from the soft-pion limits. 

\subsection{ Soft photon  constraints}
The point $s=\qdd$ is special because it corresponds to the limit of
the photon becoming soft, $q_1=0$. In this limit, we expect the helicity
amplitudes to vanish, after subtracting the Born term,  according to
the general theorem of Low~\cite{Low:1958sn}.
This follows, in the present case, simply from the general expressions
of the helicity amplitudes~\rf{helicrels} in terms of the invariant
functions. These expressions show that the amplitudes
vanish when $s=\qdd$ except if $A(s,t,\qdd)$, $B(s,t,\qdd)$ or
$C(s,t,\qdd)$ has a pole in the soft photon limit. This is the case
for the Born term parts $A^{Born}(s,t,\qdd)$, $B^{Born}(s,t,\qdd)$
see~\rf{ABborn} which have a pole when $t=\mpid$ or $u=\mpid$, and one has
\be
t-\mpid= 2q_1\cdot p_1,\ u-\mpid= 2q_1\cdot p_2
\en
which indeed vanish in the soft photon limit. The remaining parts, by
definition, have no such poles in $t$ or $u$.
The soft photon constraint implies that the subtraction functions
$a^{I}$ and $b^{I}$ in eq.~\rf{OMformule} obey the following linear relation
\be
a^I(\qdd)+\qdd b^I(\qdd)+ (\qdd)^2\,\left[
F^v_\pi(\qdd)\hat{J}^{I,\pi}(\qdd)
+\sum_{V} F_{V\pi}(\qdd) J^{I,V}(\qdd)\right]=0
\en
with
\be
\hat J^{I,\pi}(\qdd)
=\left.{\partial J^{I,\pi}(s,\qdd) \over\partial
  s}\right\vert_{s=\qdd}\ ,
\en
which we can write as
\be\lbl{Jhat}
\hat J^{I,\pi}(\qdd)=-{\sqrt{4-I}\over\sqrt3\,\pi}
\int_{4\mpid}^\infty {ds'\over (s'-\qdd)} {d\over ds'} 
\left[{\sin\delta_0^I(s')\over (s')^2 \vert\Omega_0^I(s')\vert}\,
  (4\mpid L_\pi(s')-2\qdd) \right]\ .
\en
where we have replaced  $\bar{h}^{I,Born}$ by its explicit expression
(see~\rf{pwborn0},~\rf{isoborn}) and integrated by parts.

The appearance of a derivative in the integrand of eq.~\rf{Jhat}
(which is needed when $\qdd\ne0$) may seem peculiar but  simply
results from the $(s-\qdd)$ denominator in the Born amplitude. One
immediate consequence of this structure is that the integral in
eq.~\rf{Jhat} diverges when $q^2=4\mkd$ because the phase-shift
$\delta_0^I(s')$ exhibits a cusp at the $K\Kbar$ threshold. This
problem is caused by the approximation of using one-channel Omn\`es
formulae, while the cusp is due to the opening of an inelastic
channel. We show in appendix~\sect{Nodivergence} that no divergence
arises if one 
consistently implements two-channel unitarity in the Omn\`es
method. The one-channel Omn\`es formalism should be used only for $q^2
< 1$ GeV$^2$.      
\subsection{ Adler zero, chiral expansion}
In the chiral limit,  current algebra easily shows that the
$\gamma^*(\qdd)\to \pi^0\pi^0\gamma$ amplitude $W^{\mu\nu}$ vanishes when one
of the pions becomes soft\footnote{In the case of charged pions, 
a sum rule was derived by Terazawa~\cite{Terazawa:1971fr} giving the
amplitude for two off-shell photons producing two soft pions.}.  
In this limit, e.g. $p_1\to 0$,
the tensors $T_1^{\mu\nu}$, $T_2^{\mu\nu}$,  $T_3^{\mu\nu}$ are no
longer independent. The following relations hold among them, 
\be
\left.T_2^{\mu\nu}\right\vert_{\mpi=0,p_1=0}\equiv 2\qdd T_1^{\mu\nu},\quad 
\left.T_3^{\mu\nu}\right\vert_{\mpi=0,p_1=0}\equiv-2\qdd T_1^{\mu\nu}
\en
(the sign $\equiv$ means that equality holds up to terms which vanish
when contracted with $\epsilon_1$) such that one has
\be
\left.W^{\mu\nu}\right\vert_{\mpi=0,p_1=0}=\left[ A
  +2\qdd(B-C)\right]T_1^{\mu\nu}\ .
\en
and the soft pion theorem implies
\be
\left. A +2\qdd(B-C)\right\vert_{\mpi=0,\, t=0,\ s=0}=0
\en
which holds true for any value of $\qdd$.
Let us then introduce a function of $s$ 
\be
W(s,\qdd)\equiv
\left. A(s,t,\qdd)+2\qdd(B(s,t,\qdd)-C(s,t,\qdd))
\right\vert_{t=\mpid} \ ,
\en
which, in the chiral limit,  behaves as $W(s,\qdd)\sim\lambda(\qdd) s$
at small $s$. In the physical, massive pion case,  it behaves at small
$s$ as
\be\lbl{adlerW}
W(s,\qdd)\sim (\lambda(\qdd)+a(\qdd)) s +b(\qdd)
\en
where $a,\ b$ are $O(\mpid)$. The function $W$ should 
therefore display an Adler zero as a function of $s$. The same holds
for the helicity amplitude $H^n_{++}$  when $t=\mpid$ (which
corresponds to $\cos\theta=1/\sigma_\pi(s)$), since it can be 
written as
\be\lbl{HppAdler}
\left.H^n_{++}(s,\qdd,\theta)\right\vert_{t=\mpid}={1\over2}(s-\qdd)\left[
  W(s,\qdd)-2(s-4\mpid)B(s,t=\mpid,\qdd)\right]\ 
\en
and thus can be cast into a form similar to eq.~\rf{adlerW} for small 
values\footnote{The Adler zero can disappear for exceptional values of
  $\qdd$ such that $\lambda(\qdd)-2B(0,t=\mpid,\qdd)=0$.
} of $s$. 
The value of the Adler zero $s_A$ should be small, $O(\mpid)$, and
depend on the value of $\qdd$. For illustration, in
the case of the vector exchange amplitude, $H^V_{++}$, one has 
\be\lbl{adlerHv}
\left.H^V_{++}(s_A,\qdd,\theta)\right\vert_{t=\mpid}=0,\quad
s_A={\mpid\,(2\mvd-2\mpid-\qdd)\over\mvd-2\mpid}\ . 
\en

In the chiral expansion, the amplitudes $\gamma^*\to \gamma
\pi^0\pi^0,\ \gamma \pi^+\pi^-$ have been computed at NLO in
refs.~\cite{Donoghue:1988eea,Unkmeir:1999md}. The
results, for the non-vanishing helicity amplitudes, are recalled below  
\bea\lbl{nlohelic}
&&\left.H_{++}^{n}\right\vert_{NLO}=  {2(s-\mpid) \over
  \fpid}\,\bar{\cal G}(s,\qdd) 
\nonumber\\
&&\left.H_{++}^{c}\right\vert_{NLO}=  {s\over \fpid}\bar{\cal  G}(s,\qdd)
+(\bar{l}_6-\bar{l}_5)\,{s-\qdd\over 48\pi^2\fpid} + \,H_{++}^{Born}
\nonumber\\
&&\left.H_{+-}^{c}\right\vert_{NLO}=\,H_{+-}^{Born},\ H_{+0}^{c}=\,H_{+0}^{Born} 
\ena
with
\be
\bar{\cal G}(s,\qdd)= 
{s\bar{G}_\pi(s)     -\qdd\bar{G}_\pi(\qdd)\over s-\qdd}
-\qdd {\bar{J}_\pi(s)-\bar{J}_\pi(\qdd)    \over s-\qdd}
\en
and using the definitions of ref.~\cite{bellucci} for the loop functions
$\bar{J}_\pi$ and $\bar{G}_\pi$,
\bea\lbl{Jbar}
&&\bar{J}_\pi(z)={1\over16\pi^2}
\left(2+\betapi(z)\log{\betapi(z)-1\over\betapi(z)+1}\right),\ 
\nonumber\\
&&\bar{G}_\pi(z)=-{1\over16\pi^2}
\left(
1+{\mpid\over z}\log^2{\betapi(z)-1\over\betapi(z)+1}\right)\ .
\ena
These functions satisfy the relation
\be
{d\over dz}\left( z \bar{G}_\pi(z)\right)=z {d\over dz} \bar{J}_\pi(z)
\en
which ensure that $H^n_{++}$ and $H^c_{++}-H^{Born}_{++}$ vanish at
the soft photon point $s=\qdd$.
In eqs.~\rf{nlohelic} the NLO expression for the pion form factor 
which enters in $H^{Born}$  must be used i.e.~\cite{gl84}
\be
\left. F^v_\pi(\qdd)\right\vert_{NLO}=1 
+{1\over6\fpid}(\qdd-4\mpid)\bar{J}_\pi(\qdd)+{\qdd\over96\pi^2\fpid}
\left(\bar{l}_6- {1\over3}\right)\ .
\en
The NLO amplitude $\left.H_{++}^n(s)\right\vert_{NLO}$ has an Adler
zero at $s=\mpid$ which does not depend on the value of $\qdd$.

\subsection{Dispersive amplitudes with chiral constraints}\lblsec{chiralcons}
In order to  implement  chiral constraints in our dispersive
representation of the amplitude in a transparent way, we redefine the
subtraction functions $b^I(\qdd)$ such that all the integral pieces
are multiplied by a factor of $s$: 
\bea\lbl{completehelic}
&&H^I_{++}(s,\qdd,\theta)= 
F^v_\pi(\qdd) \bar{H}^{I,Born}_{++}(s,\qdd,\theta)
+\sum_{V=\rho,\omega}F_{V\pi}(\qdd) \bar{H}^{I,V}_{++}(s,\qdd,\theta) 
\nonumber \\
&&+\Omega_0^I(s) \Bigg\{ (s-\qdd)\, b^I(\qdd)
+s F^v_\pi(\qdd)\left[ {s (J^{I,\pi}(s,\qdd)- J^{I,\pi}(\qdd,\qdd))\over s-\qdd }
-\qdd\hat{J}^{I,\pi}(\qdd)\right]\nonumber \\
&& +s \sum_{V=\rho,\omega} F_{V\pi}(\qdd) \Big[ s J^{I,V}(s,\qdd)-\qdd
  J^{I,V}(\qdd,\qdd)\Big]\Bigg\}\ .
\ena 
The value at $s=0$ of the $\pi^0\pi^0$ amplitude is now given simply
by
\be\lbl{completehelic0}
H^n_{++}(0,\qdd,\theta)=\sum_{V=\rho^0,\omega} H^{n,V}_{++}(0,\qdd,\theta)-\qdd b^n(\qdd)
\en
where $b^n(\qdd)$ (and similarly $b^c(\qdd)$) is given from
eq.~\rf{Hisomatrix} in terms of $b^I(\qdd)$
\be\lbl{bisomatrix}
\left(\ba{c}
\sqrt2\, b^c(\qdd)\\
b^n(\qdd)\\
\ea\right)=\mathbf{C}
\left(\ba{c}
b^0(\qdd)\\
b^2(\qdd)\\
\ea\right)\ .
\en
Consistency with the soft pion theorem requires that the right-hand
side of eq.~\rf{completehelic0} should vanish in the chiral limit,
i.e. $b^n(\qdd)\sim O(\mpid)$, at least when $\qdd\ne0$. The chiral behaviour
when $\qdd=0$ is actually different. This can be seen by comparing with
the NLO chiral amplitude at $s=0$ and $q^2 << \mpid$, 
\be
H^{n,NLO}_{++}(s=0,\qdd,\theta)={-2\mpid\over\fpid}(\bar{G}_\pi(\qdd)
-\bar{J}_\pi(\qdd))={\qdd\over 96\pi^2\fpid}\left(1
+{\qdd\over15\mpid}+\cdots\right) 
\en
from which one deduces that  $b^n(0)=-1/96\pi^2\fpid$ in the chiral limit.

%%%%%%%%%%%%%%%%% Dispersive versus chiral figure
\begin{figure}[ht]
\bc
\includegraphics[width=0.6\linewidth]{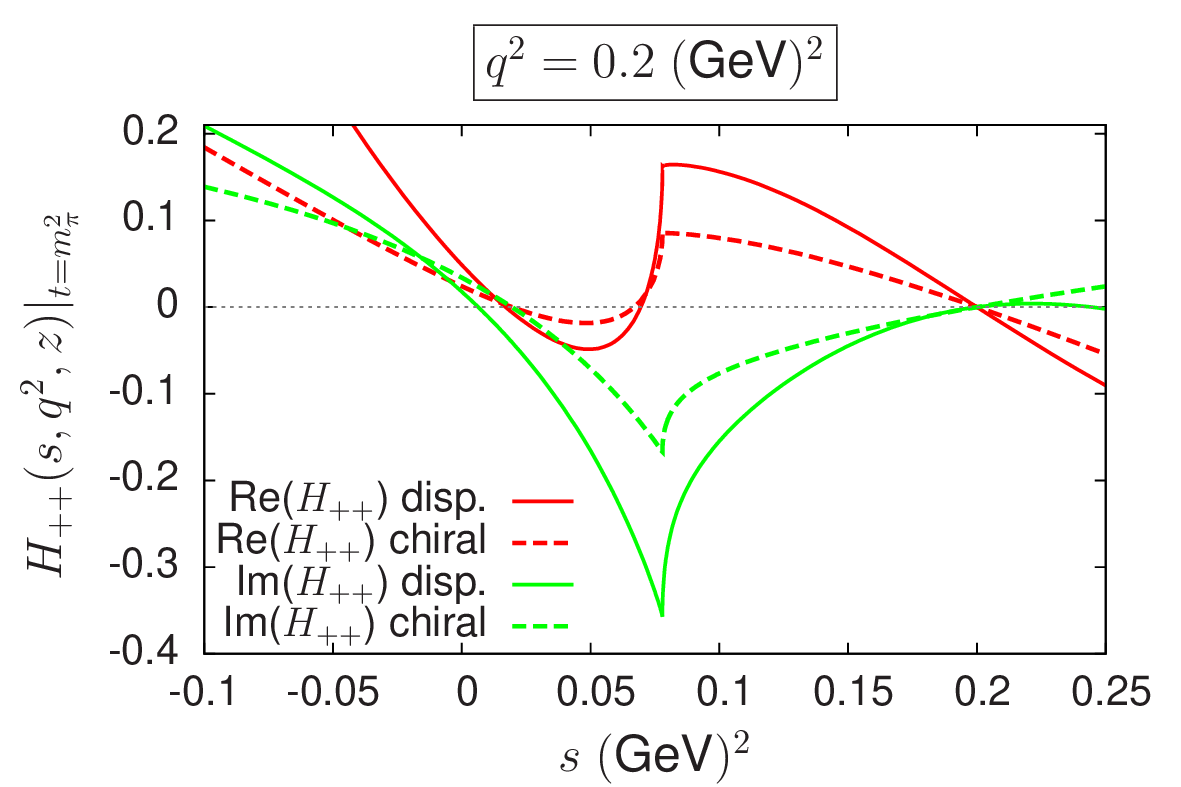}
\includegraphics[width=0.6\linewidth]{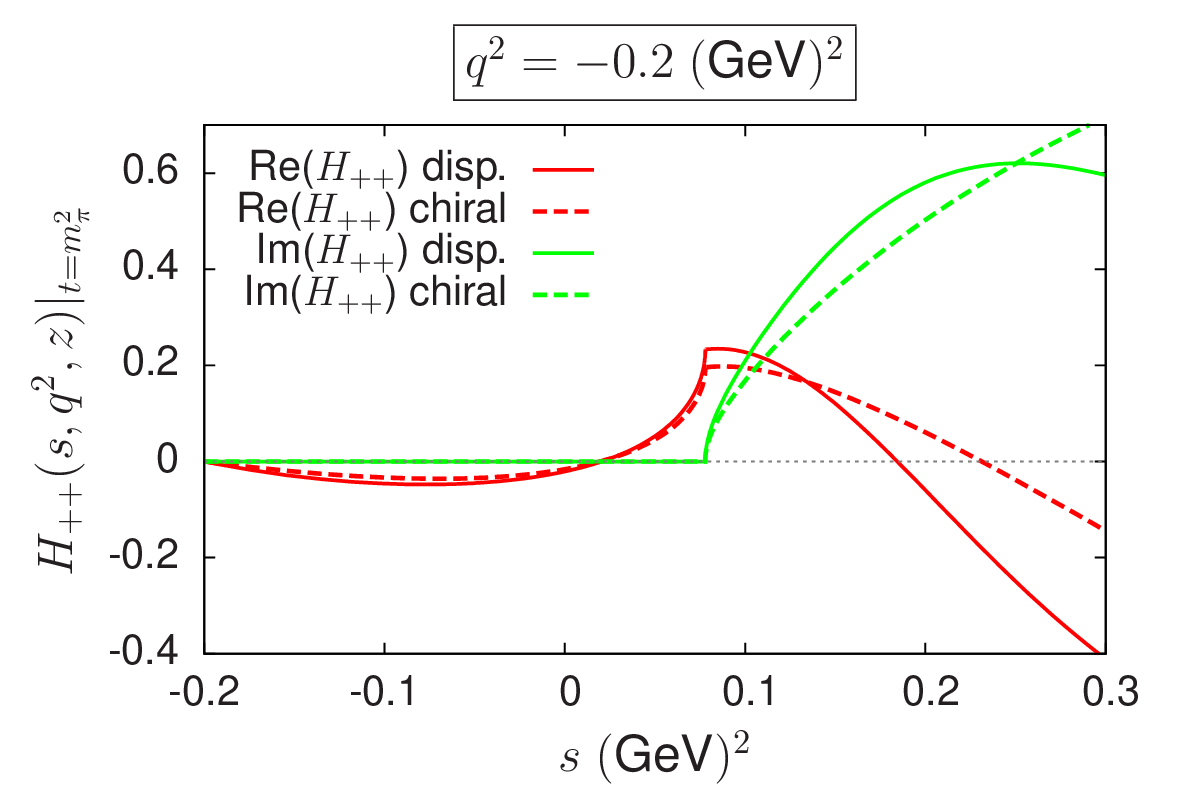}
\caption{Illustration of the Adler and soft photon zeros in the
  dispersive amplitude $H^n_{++}$ and in the corresponding NLO chiral
  approximation. The upper plot corresponds to $q^2=0.2$ GeV$^2$ and the
  lower one to $q^2=-0.2$ GeV$^2$.}
\lblfig{compchiral}
\ec
\end{figure}
%%%%%%%%%%%%%%%%% Dispersive versus chiral figure
As will be discussed in the next section, we 
assume the subtraction functions $b^I(\qdd)$ to be slowly varying,
except near the positions of the vector resonances, and we will
constrain them from experimental data at $\qdd=0$ and $\qdd > 4\mpid$. 
The Adler and soft photon zeros in the resulting amplitudes are
illustrated in fig.~\fig{compchiral}. The figure shows $H_{++}$ with
$t=\mpid$,  as a  function of $s$ in the case of  both positive and
negative values of $q^2$. In the latter case, the NLO chiral and the
dispersive amplitudes are very close in the region $q^2 \le s \le
4\mpi^2$. The amplitude is also rather small in this region because of
the two zeros. In the case of positive $q^2$, there is a visible
difference between the two amplitudes. To a large extent, this
reflects the influence of the pion form factor, which grows rapidly
for positive $q^2$, and is set equal to 1 in the NLO amplitude. The
figure also shows that both the real and imaginary parts of
the dispersive amplitude  display an Adler zero, but they do not
coincide as in the NLO case. Furthermore, their location varies a
function of $q^2$.

\subsection{Comparison with some other approaches}
Eq.~\rf{completehelic} represents our final result for the dispersive
representation of the $\gamma^*\gamma \to \pi\pi$ or $\gamma^* \to
\gamma\pi\pi$ amplitudes. There has been a long lasting interest in
the literature for  the closely related amplitudes describing the
decays of the $\rho$, $\omega$ mesons into $\gamma\pi\pi$. An
illustrative list of references is~\cite{Singer:1962qr,Singer:1963ae,Renard:1969si,Fajfer:1990cd,Bramon:1992kr,Huber:1995np,Marco:1999df,Guetta:2000ra,Bramon:2001un,Palomar:2001vg,Gokalp:2003uf,Escribano:2006mb,Eidelman:2010ta,Achasov:2010fh}. Much
of the previous work is based on computing the amplitudes from chiral
Lagrangians which also include a few light resonances. Chiral
Lagrangians automatically enforce chiral as well as QED Ward
identities. Furthermore, Feynman diagrams satisfy analyticity
properties, as well as unitarity relations if loops are computed. All
results, therefore, could be written in a way formally analogous to
eq.~\rf{completehelic}. The Born term rescattering piece, for
instance, would correspond to the pion loop contribution in a Lagrangian
calculation.  The amplitude~\rf{completehelic} further includes the
rescattering contributions associated with the $\rho$ and $\omega$ exchange
amplitudes,  which would correspond to $\pi+\rho$ and $\pi+\omega$
loops in a Lagrangian approach (see appendix~\sect{Disptriangles}). Such
contributions seem not to have 
been considered previously. Additionally, elastic unitarity for
the $\pi\pi\to\pi\pi$ partial-wave scattering amplitudes is enforced
exactly in the dispersive expression~\rf{completehelic}. This property
is also correctly satisfied in the unitarized ChPT
approaches~\cite{Marco:1999df,Palomar:2001vg}, but not in the
resonance Lagrangian ones. Finally, the rescattering contribution from
the $I=2$ $\pi\pi$ amplitude has usually been neglected in previous work.
\section{Comparison with experiment}\lblsec{compexp}
\subsection{$\pi\pi$ phase-shifts and Omn\`es functions}
The expressions for the $\gamma\gamma^*$ amplitudes involve the Omn\`es
functions $\Omega_0^I(s)$ constructed from the $I=0,2$ $S$-wave $\pi\pi$
phase-shifts $\delta_0^I$, 
%modif here: explicit omnes formula
\be\lbl{omnesrecall}
\Omega_0^I(s)=\exp\left[{s\over\pi}\int_{4\mpid}^\infty ds'
  {\delta_0^I(s')\over s'(s'-s)}
\right]\ .
\en
In using one-channel Omn\`es functions one ignores
inelastic channels in the unitarity relations. The $I=0$ channel is
peculiar in this respect, because the inelasticity associated with the
$K\Kbar$ channel sets in rather sharply as a effect of the $f_0(980)$
resonance. This resonance also causes the phase-shift $\delta_0^0$ to
raise very  rapidly at 1 GeV, which  gives rise to a large peak in
the Omn\`es function $\Omega_0^0$. As has been pointed out in
ref.~\cite{Oller:2008kf}, it is useful to make use of
experimental information at the $f_0(980)$ peak even if the formulae
are to be applied at smaller energies. The $f_0(980)$ peak in the case of the
$\gamma\gamma\to\pi\pi$ amplitudes is observed to be rather
small. In fact, the peak was clearly observed only recently by the Belle
collaboration $\gamma\gamma$
experiments~\cite{bellemori2pic_1,belleuehara2pi0_1}. This implies
that the polynomial parameters in the $\gamma\gamma$ amplitude must be
such as to produce a zero close to one GeV in the coefficient of the
Omn\`es function. An equivalent method for generating a zero is
to make use of a modified Omn\`es function $\Omega[\phi_0^0]$, constructed 
with a phase $\phi_0^0$ which satisfies\footnote{
Alternatively, ref.~\cite{Hoferichter:2011wk} discuss the idea of using an
Omn\`es function with a cutoff, i.e. setting $\phi_0^0=0$ for $s\ge
\Lambda$. In the present context, the divergence of the function
$\hat{J}^{0,\pi}(\qdd)$ leads to a reduced range of applicability, as
a function of $\qdd$, as compared to the prescription of eq.~\rf{phi00}.} 
\be\lbl{phi00}
\ba{ll}
\phi_0^0(s)=\delta_0^0(s),   & s \le s_\pi\\
\phi_0^0(s)=\delta_0^0(s)-\pi,& s >   s_\pi\\
\ea
\en
and $s_\pi$ is such that $\delta_0^0(s_\pi)=\pi$, which is a point
close to the $K\bar{K}$ threshold. The modified and original functions
satisfy the simple relation $\Omega[\phi_0^0]=(1-s/s_\pi)
\Omega[\delta_0^0]$.

We used phase-shifts $\delta_0^0$, $\delta_0^2$
based, at low energies, on the twice-subtracted Roy equations analysis
from ref.~\cite{Ananthanarayan:2000ht}. The two scattering lengths $a_0^0$,
$a_0^2$ have been updated to the values given  by the NA48/2
collaboration~\cite{Batley:2010zza}. In the case of $\delta_0^0$, the
Roy solutions are extended in energy up to the $K\Kbar$ threshold (see
~\cite{Moussallam:2011zg}). Above 1 GeV, the phase-shifts are taken
from fits to experimental data. 

\subsection{Case $q^2=0$}
Setting $q^2=0$, the amplitudes $H_{\lambda\lambda'}$ correspond to
photon-photon scattering, $\gamma\gamma\to\pi\pi$. In this case, the
helicity amplitude $H_{+0}$ vanishes identically and the 
differential cross-section is given by 
\be
{d\sigma\over d\cos\theta}= {\alpha^2 \pi \over 4 s }\betapi(s)
\left( \vert H_{++}\vert^2 + \vert H_{+-}\vert^2 \right)
\en
The values of the subtraction functions at $q^2=0$ can be related
to the pion electric and magnetic polarizabilities $\alpha_\pi$,
$\beta_\pi$. These two observables indeed parametrize the pion Compton
scattering amplitude near threshold (see e.g.~\cite{Donoghue:1993kw}
and references therein). One can then relate the polarizability
difference $\alpha_\pi-\beta_\pi$ to the helicity amplitude $H_{++}$
computed in the limit $t\to \mpid$, $s\to 0$ after subtracting the Born
amplitude 
\bea\lbl{polarH++}
&& \mpi(\alpha_{\pi^0}-\beta_{\pi^0})= 
\lim_{s\to0,\,t\to\mpid}{2\alpha\over  s}H^n_{++}(s,q^2=0,\theta) \nonumber\\
&& \mpi(\alpha_{\pi^+}-\beta_{\pi^+})=
\lim_{s\to0,\,t\to\mpid}{2\alpha\over  s}\hat{H}^c_{++}(s,q^2=0,\theta)
\ena
where, in the charged case, 
\be\lbl{Hhatdef}
\hat{H}^c_{\lambda\lambda'}\equiv
H^c_{\lambda\lambda'}-H^{Born}_{\lambda\lambda'}\ .
\en
In the approach followed here, the following simple relation thus
holds between the polarizabilities and the values of the
subtraction functions at $\qdd=0$ 
\bea
&& (\alpha_{\pi^0}-\beta_{\pi^0})= {2\alpha\over\mpi}\left( b^n(0)
-{4\mpid}\tilde C_{\rho^0}\,\widetilde{BW}_\rho(\mpid)
-{4\mpid\,\tilde{C}_{\omega}\over m_\omega^2-\mpid} \right)
\nonumber\\
&& (\alpha_{\pi^+}-\beta_{\pi^+})={2\alpha\over\mpi}\left( b^c(0)
-{4\mpid}\tilde C_{\rho^+}\,\widetilde{BW}_\rho(\mpid)  \right)\ .
\ena
At present, the values of the pion polarizabilities cannot be
considered as precisely determined experimentally. It was observed in
ref.~\cite{Babusci:1991sk} that NLO ChPT predictions were in
qualitative agreement with the available $\gamma\gamma\to\pi\pi$
cross-sections. Their discussion was improved in
ref.~\cite{Donoghue:1993kw} who combined ChPT with Omn\`es dispersive
representations. New measurements of $\gamma\gamma\to \pi^0\pi^0$,
$\pi^+\pi^-$ covering the very low energy region are planned at
KLOE~\cite{AmelinoCamelia:2010me}. For charged pions, a new experiment
is under way at COMPASS~\cite{Guskov:2010zz} which aims at measuring
the Compton amplitude and the polarizability by the Primakov method.
There have also been attempts to determine the polarizabilities from
unsubtracted dispersion relations leading, however, to somewhat
conflicting results~\cite{Filkov:1982cx,Pasquini:2008ep}.
The result for the polarizability difference in the chiral expansion at
NLO is easily obtained from eqs.~\rf{nlohelic},
\be
\left.(\alpha_\piz-\beta_\piz)\right\vert_{NLO}\simeq
-1.0\cdot10^{-4}\ \hbox{fm}^3,\quad 
\left.(\alpha_\pip-\beta_\pip)\right\vert_{NLO}\simeq
 6.0\cdot10^{-4}\ \hbox{fm}^3\ 
\en
(with $F_\pi=92.2$ MeV, $\bar{l}_6-\bar{l}_5=3.0$). The calculation of
the $\gamma\gamma$ amplitudes at NNLO have been
performed~\cite{bellucci,burgi,Gasser:2005ud,Gasser:2006qa}. However,
quantitative results for the polarizabilities at NNLO are affected by
an uncertainty due to the fact that the $O(p^6)$  chiral coupling constants
are not known at present.   
For definiteness, we will use here 
the estimates obtained in ref.~\cite{garcia-martin} from a coupled
channel MO treatment of  the two sets of measurements by the Belle
collaboration~\cite{bellemori2pic_1,belleuehara2pi0_1}. These data
have very high statistics but do not cover the very low energy
region. This analysis favoured the following value for the neutral
pion polarizability difference\footnote{ 
In the fit, the dipole and quadrupole polarizabilities of the $\pi^0$
were allowed to vary  subject to the constraint that the combination
$6(\alpha_\piz-\beta_\piz)_{dipole}+\mpid(\alpha_\piz-\beta_\piz)_{quadrupole}$
is given by a chiral sum rule.}, 
\be\lbl{polarexpn}
(\alpha_{\pi^0}-\beta_{\pi^0})=-(1.25\pm0.16)\,10^{-4}\ {\hbox{fm}^3}
\en
while the charged polarizability difference was constrained to lie in
the range predicted by the two-loop calculation plus resonance
modelling of the LEC's performed in ref.~\cite{Gasser:2006qa}. The
data favoured values in the lower part of that range
\be\lbl{polarexpc}
(\alpha_{\pi^+}-\beta_{\pi^+})\simeq 4.7\cdot10^{-4}\ {\hbox{fm}^3}\ .
\en
Using the determinations~\rf{polarexpn},~\rf{polarexpc} 
for the couplings $\tilde{C}_V$ gives the following values for the
subtractions functions at $\qdd=0$
\be\lbl{bI(0)}
b^0(0)=-(0.66\pm0.20),\ \hbox{GeV}^{-2}\quad\
b^2(0)=-(0.54\pm0.14)          \quad \hbox{GeV}^{-2}
\en
(we have ascribed an error $\pm1.4\,10^{-4}$ to the charged
polarizabilities difference). 

\begin{figure}[hbt]
\bc
%%%%%%%%%%%%%%%%gamma-gammma--> pi^0 pi^0 figure
\includegraphics[width=0.8\linewidth]{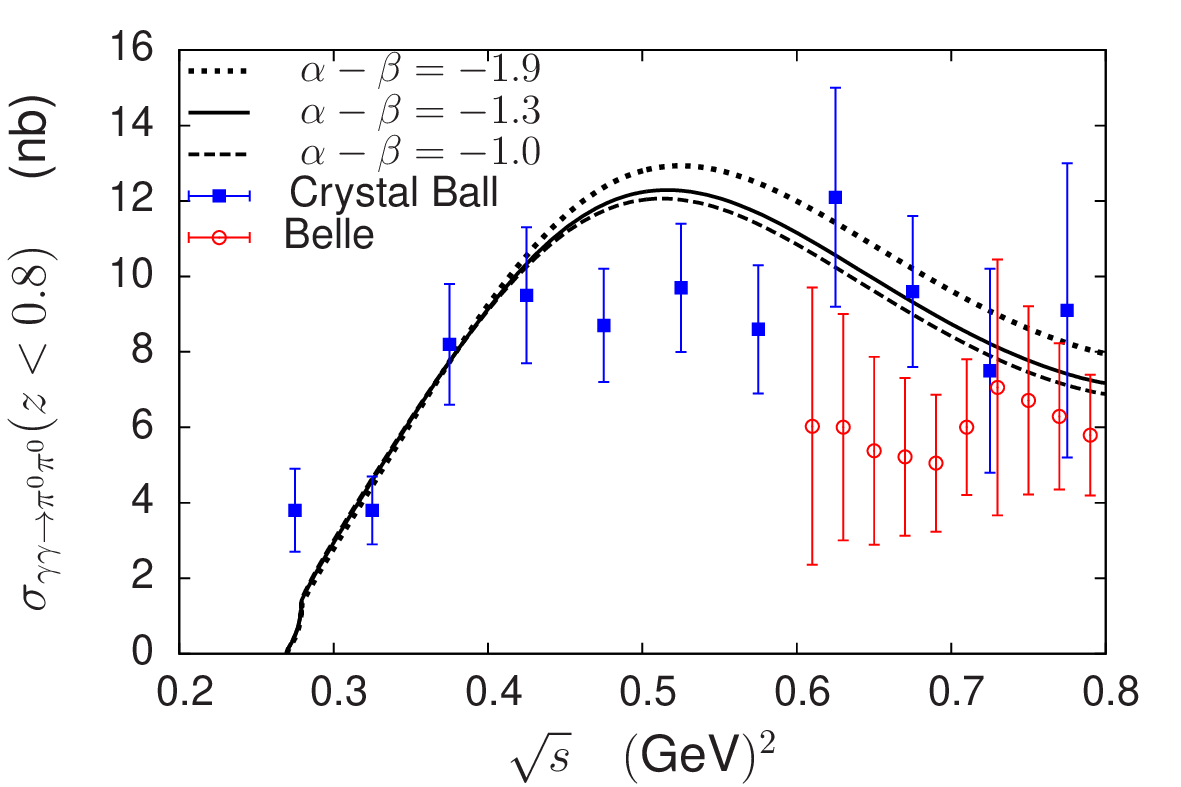}
\caption{Comparison of the
$\gamma\gamma\to\pi^0\pi^0$ cross-sections using the amplitude
$H^n_{++}$ as derived from eq.~\rf{completehelic} and
  $H^n_{+-}=H_{+-}^{n,V}$ with experiment.
The influence of varying the polarizability difference
$\alpha_{\pi^0}-\beta_{\pi^0}$ is shown.} 
\label{fig:crosssecgg}
\ec
\end{figure}

The  result for the
$\gamma\gamma\to \pi^0\pi^0$ cross-section derived from our amplitudes
using the values~\rf{bI(0)} for $b^I(0)$
is shown on fig.~\fig{crosssecgg} and compared to the experimental
measurements from refs.~\cite{crystalball,belleuehara2pi0_1}. Note
that the cross-section displays a cusp at $\sqrt{s}=4m^2_{\pip}$, due
to the $\pi^0-\pi^+$ mass difference, which was discussed in
ref.~\cite{Kaiser:2011zz} using ChPT.
%%%%%%%%%%%%%%%%gamma-gammma--> pi^0 pi^0 figure

\subsection{Case $\qdd\ne  0$: $F^v_\pi$, $F_{\omega\pi}$,
  $F_{\rho\pi}$ form factors}
In order to address the case with $q^2\ne 0$ we must specify the $q^2$
dependence of the three form factors which enter into the expression
of the amplitude~\rf{completehelic}. They were defined from the
relevant matrix elements of the electromagnetic current operator by
eqs.~\rf{ffactordef},~\rf{Vpiffdef}. We will employ usual
phenomenological descriptions based on superposition of Breit-Wigner
type functions associated with the light vector resonances. 
We give some details on these in appendix~\sect{Formfactors}. The pion
form factor, of course, is known rather precisely from experiment. Some
experimental data exist also for the $\omega\pi$ form factor in two
kinematical regions surrounding the peak of the $\rho$ meson. The data
in these two ranges are compatible with the simple model used except,
possibly, in a small energy region (see appendix~\sect{Formfactors}
for more details). The $F_{\rho\pi}$ form factor, finally, is more
difficult to isolate experimentally than $F_{\omega\pi}$, because of
the width of the $\rho$. We used the same type of modelling together
with symmetry arguments to fix the parameters.

\subsection{Case $\qdd\ne  0$: subtraction functions}
The values of $b^0(\qdd)$, $b^2(\qdd)$ when $\qdd\ne0$
are a priori not known and must thus be determined from experiment. Given
detailed experimental  data on $e^+ e^-\to \gamma\pi^0\pi^0$ and 
$e^+ e^-\to\gamma^*\to \gamma\pi^+\pi^-$, one
could determine these functions for each $q^2$ by performing a fit of
the differential $d\sigma/ds$ cross sections. In practice, one expects
that a simple parametrization of $q^2$ dependence should be
adequate. We adopted the following form,  which involves two arbitrary
parameters,  
\be\lbl{biq2dep}
\ba{ll}
b^n(\qdd)=b^n(0) F(\qdd) \!\!\!\!& +\beta_\rho
(GS_\rho(\qdd)-1)+\beta_\omega(BW_\omega(\qdd)-1) \\
b^c(\qdd)=b^c(0)& +\beta_\rho
(GS_\rho(\qdd)-1)+\beta_\omega(BW_\omega(\qdd)-1) \\
\ea
\en
with
\be
F(\qdd)=192\pi^2\,{ \mpid(\bar{J}_\pi(\qdd)-\bar{G}_\pi(\qdd))\over\qdd}\ .
\en
The relation between $b^0$, $b^2$ and $b^n$, $b^c$ is given in
eq.~\rf{bisomatrix}. This form~\rf{biq2dep} is  motivated by the
discussion concerning the chiral limit. Assuming that the 
parameters $\beta_\rho$, $\beta_\omega$ are $O(\mpid)$ ensures that
$b^n(\qdd)$, $b^c(\qdd)$ have the correct chiral limit behaviour at
$\qdd\ne0$ as well as $\qdd=0$ (see sec.~\sect{chiralcons}).

%table result of fit: Fits  28 Mars 2013
\begin{table}[ht]
\bc
\bt{r|c|c|l}\hline\hline
\B $\beta_\rho$ & $\beta_\omega$ & $\chi^2/N_{dof}$ & ref.\\ \hline
\TT$ 0.14\pm0.12$ & $(-0.39\pm0.12)\,10^{-1}$ & $20.2/27 $ &
Achasov~\cite{Achasov:2002jv}\\ 
   $-0.13\pm0.15$  &$(-0.31\pm0.15)\,10^{-1}$&  $15.0/21 $ & 
Akhmetshin~\cite{Akhmetshin:2003rg}\\
   $ 0.05\pm0.09$ & $(-0.37\pm0.09)\,10^{-1}$ & $38.1/50  $ & Combined \\
\hline\hline
\et
\caption{\sl Results of fitting the two-parameter dependence of the
  subtraction functions ( see eq.~\rf{biq2dep}) to the
  experimental data. }
\lbltab{fitresults}
\ec
\end{table}
%modif here
We consider the experimental data in the region $\sqrt{s}\le
0.95$ GeV where it is an acceptable approximation to ignore
the effect of inelasticity in $\pi\pi$ scattering. We also ignore the
effect of $\pi\pi$ rescattering in $D$ or higher partial-waves, since
the corresponding $\pi\pi$ phase-shifts are small in this
region. Note, however, that $J\ge 2$ partial-waves in the 
$\gamma^*\to\gamma\pi\pi$ amplitudes are not necessarily small, 
except very close to the $\pi\pi$ threshold.
They are included via $H_{\lambda\lambda'}^{Born}$ (for charged
amplitudes) and $H_{\lambda\lambda'}^V$.
The results of performing fits to the data of
refs.~\cite{Achasov:2002jv} and~\cite{Akhmetshin:2003rg}
are shown in table~\Table{fitresults} and illustrated in
fig.~\fig{resultsig}. The calculation of the $\chi^2$ with
asymmetric errors is done following the prescription of the introduction
chapter of the PDG. The few data points from
ref.~\cite{Akhmetshin:2003rg} which are given as upper bounds are not
included in the fit. The figure shows the result of the combined fit
compared separately with the data of
refs.~\cite{Achasov:2002jv,Akhmetshin:2003rg}. We also show the result
obtained upon setting the two parameters $\beta_\rho$, $\beta_\omega$
to zero. The amplitude with $\beta_\rho=\beta_\omega=0$ agrees with
experiment except at the $\omega$ peak. The data essentially require
one parameter, $\beta_\omega$, to be different from zero. 

%%%%%%%%%%%%%%%%%%%%%%%figure integrated cross-sections sigma(q^2)
\begin{figure}[ht]
\bc
\includegraphics[width=0.8\linewidth]{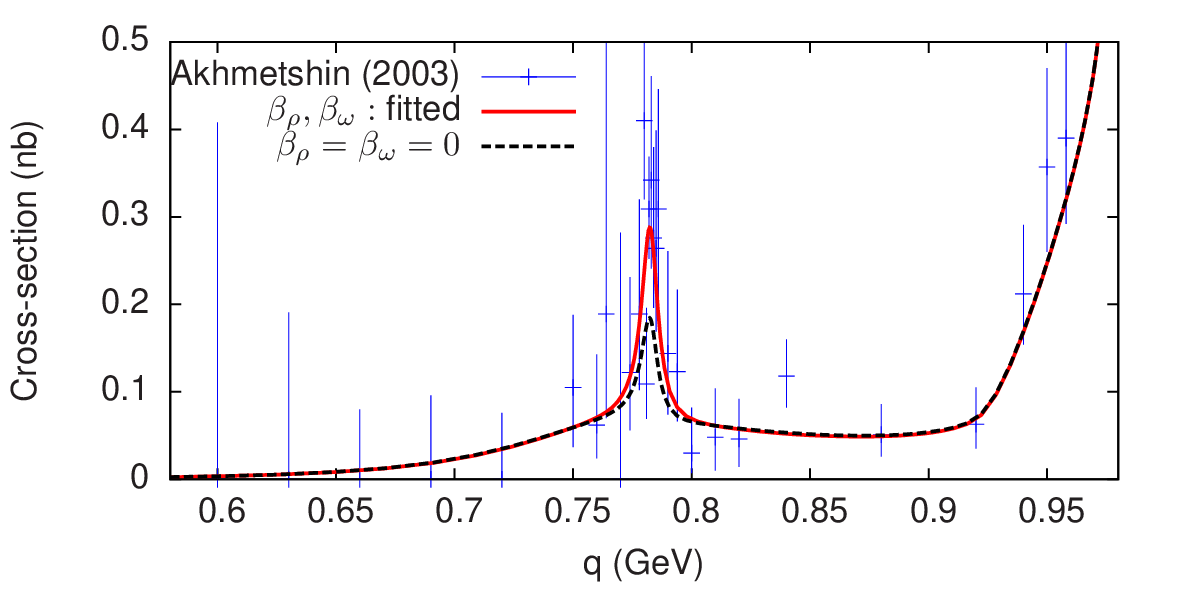}
\includegraphics[width=0.8\linewidth]{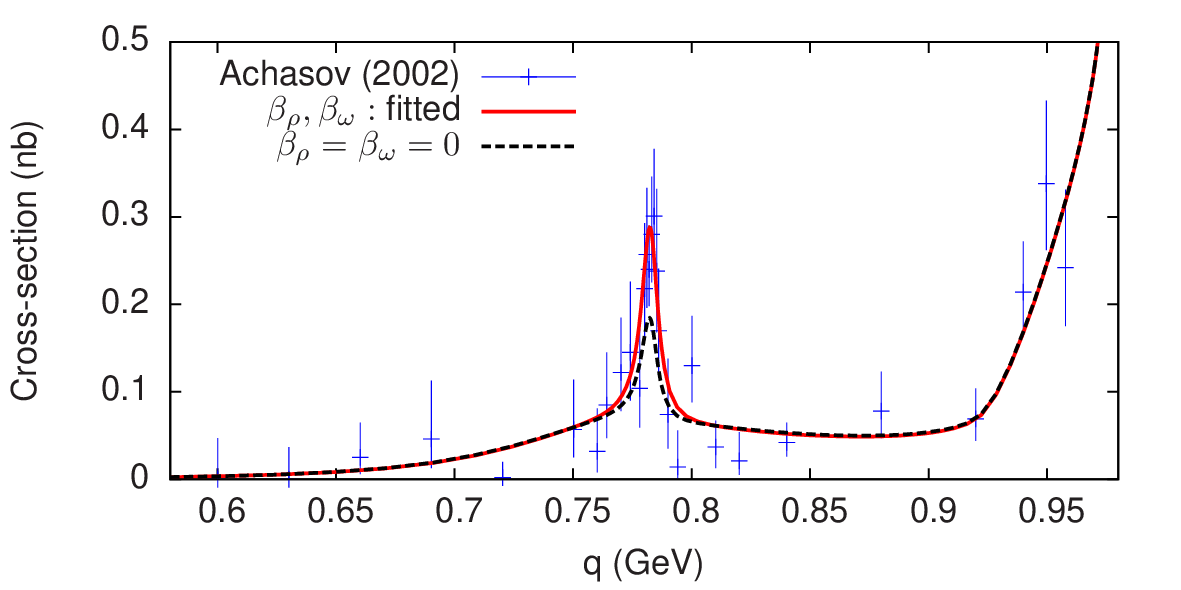}
\caption{\sl Integrated cross-section for $e^+e^-\to \gamma\pi^0\pi^0$.
The experimental results are from
refs.~\cite{Achasov:2002jv,Akhmetshin:2003rg}. The solid line is the
result of the calculation from the dispersive
representations~\rf{completehelic} with $b^I(q^2)$ parametrized as in
eq.~\rf{biq2dep} and  with  values of the parameters
$\beta_\rho$, $\beta_\omega$ obtained from a combined fit to the two
sets of experimental data (third line in table~\Table{fitresults}). The
dotted line corresponds to $\beta_\rho=\beta_\omega=0$.}  
\lblfig{resultsig}
\ec
\end{figure}
The behaviour of the differential cross-section, $d\sigma/d\sqrt{s}$ 
is shown in fig.~\fig{crosssecss} for
several values of $\qdd$. A change in the shape occurs when $\qdd\ge
(m_\omega+m_\pi)^2$. This corresponds to the appearance of the $\omega$
meson inside the Dalitz plot, which gives rise to peaks when
$t=\momegd$ or $u=\momegd$. The $\omega\pi$ threshold effect is also
clearly visible in the integrated cross-section $\sigma(\qdd)$ in
fig.~\fig{resultsig}. %modif ici % 
%%%%%%%%%%%%%%figure dsigma)s,q^2)/ds
\begin{figure}[hbt]
\bc
\includegraphics[width=0.8\linewidth]{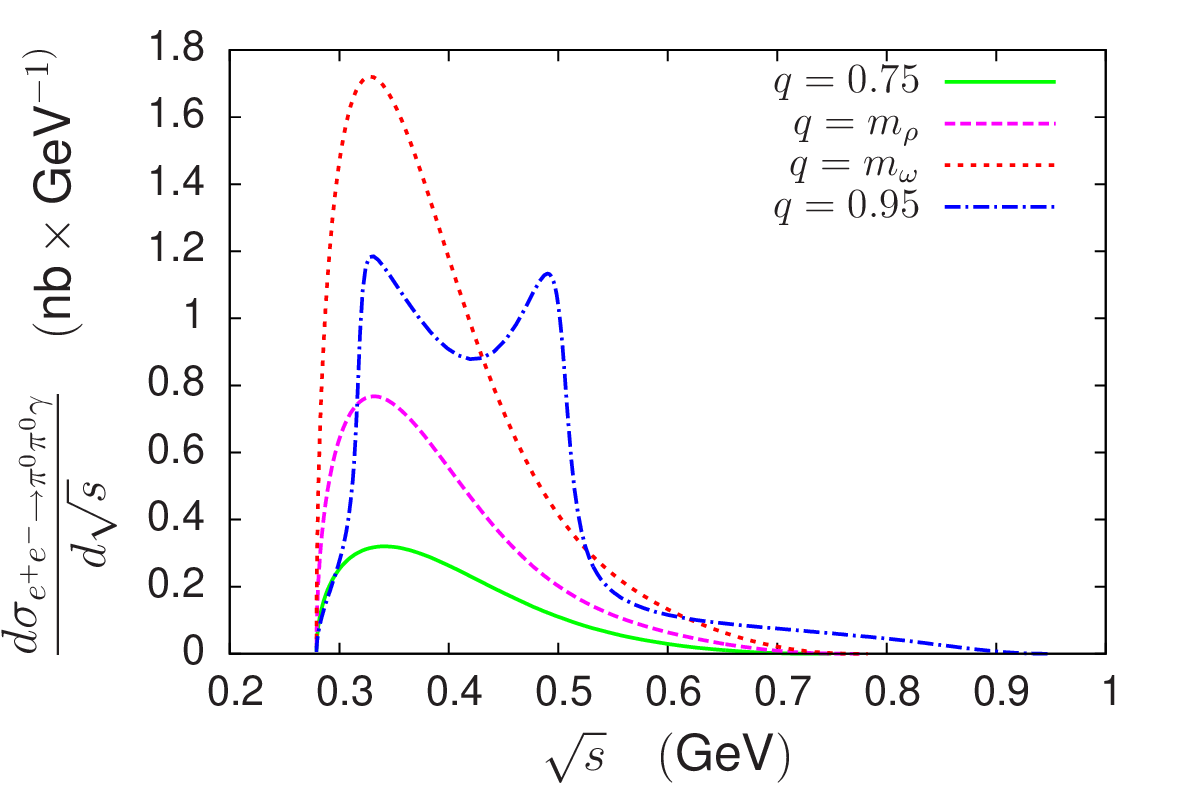}
\caption{\sl Differential cross-sections, as a function of the
  $\pi\pi$ energy $\sqrt{s}$, for various values of the virtual photon
  energy $\qdd$.}
\lblfig{crosssecss}
\ec
\end{figure}
%%%%%%%%%%%%%%figure dsigma)s,q^2)/ds

\section{Some applications}\lblsec{applications}

\subsection{Decays of the $\rho$, $\omega$  mesons into $\gamma\pi\pi$}
One can define the decay amplitude of a vector meson 
from the $\gamma^*(\qdd)\to\gamma\pi\pi$ helicity  amplitudes
when $\qdd$ is close to a resonance peak. First, one defines the
coupling $F_V$ of a vector meson to the electromagnetic current from
the matrix element 
\be
\braque{0\vert j_\mu(0)\vert V(\lambda)}= e\,m_V F_V
\epsilon_\mu(\lambda)\ .
\en
This definition is usual although it is  well defined, strictly
speaking, in the limit of a stable meson. From the quoted
values~\cite{Beringer:1900zz} of the meson decays $\rho,\omega \to e^+
e^-$ widths one obtains 
\be
F_\rho=156.5\pm 0.7,\quad  F_\omega=45.9\pm0.8 \quad \hbox{MeV}
\en
In this same zero width limit,
the amplitude which describes the vector meson decay $V\to
\gamma\pi\pi$ is related to the residue of the $V$ meson
pole in the $\gamma^*\to\gamma\pi\pi$ helicity amplitude by the LSZ
formula 
\be
T^V_{\lambda\lambda'}(s,\theta)=\lim_{\qdd=\mvd} {\qdd-\mvd\over m_V F_V} \,
e H_{\lambda\lambda'}(s,\qdd,\theta)\ .
\en
In the finite width case, the pole is replaced by a Breit-Wigner type
function in our representations and we approximate the residue by the
coefficient of this function. These Breit-Wigner type functions are
present in the parametrizations of the three form factors $F^v_\pi$,
$F_{\omega\pi}$, $F_{\rho\pi}$ (see appendix~\sect{Formfactors}, note
that $\omega-\rho$ mixing is accounted for) and also in the
subtraction functions $b^0(\qdd)$, $b^2(\qdd)$ . The differential
decay width is given in terms of  the amplitude $T^V_{\lambda\lambda'}$ by  
\be\lbl{diffwidth}
{d^2\Gamma\over ds\,d\cos\theta}=
{\alpha (\mvd-s)\over12(4\pi)^2 m_V^3} \betapi(s)
\left( \vert  T^V_{++}  \vert^2 +\vert  T^V_{+0}  \vert^2 +\vert
T^V_{+-}  \vert^2 \right)\ .
\en
From this,  and using the fitted parameters from the last line of 
table~\Table{fitresults}, we deduce the following results for the branching
fractions
\be
\ba{l}
BF(\omega\to\pi^0\pi^0\gamma)=(5.61\pm1.70)\,10^{-5},\\
BF(\rho\to \pi^0\pi^0\gamma)=(4.21\pm0.60)\,10^{-5}\ .
\ea
\en 
The result for the $\omega$ differs somewhat from that derived by the
experimental groups \cite{Achasov:2002jv,Akhmetshin:2003rg} (from the
same data), being  smaller by nearly one sigma. This illustrates that
these branching fractions are not directly measurable quantities,
unlike the $e^+e^-$ cross-sections. 

\begin{figure}[ht]
\bc
\includegraphics[width=0.7\linewidth]{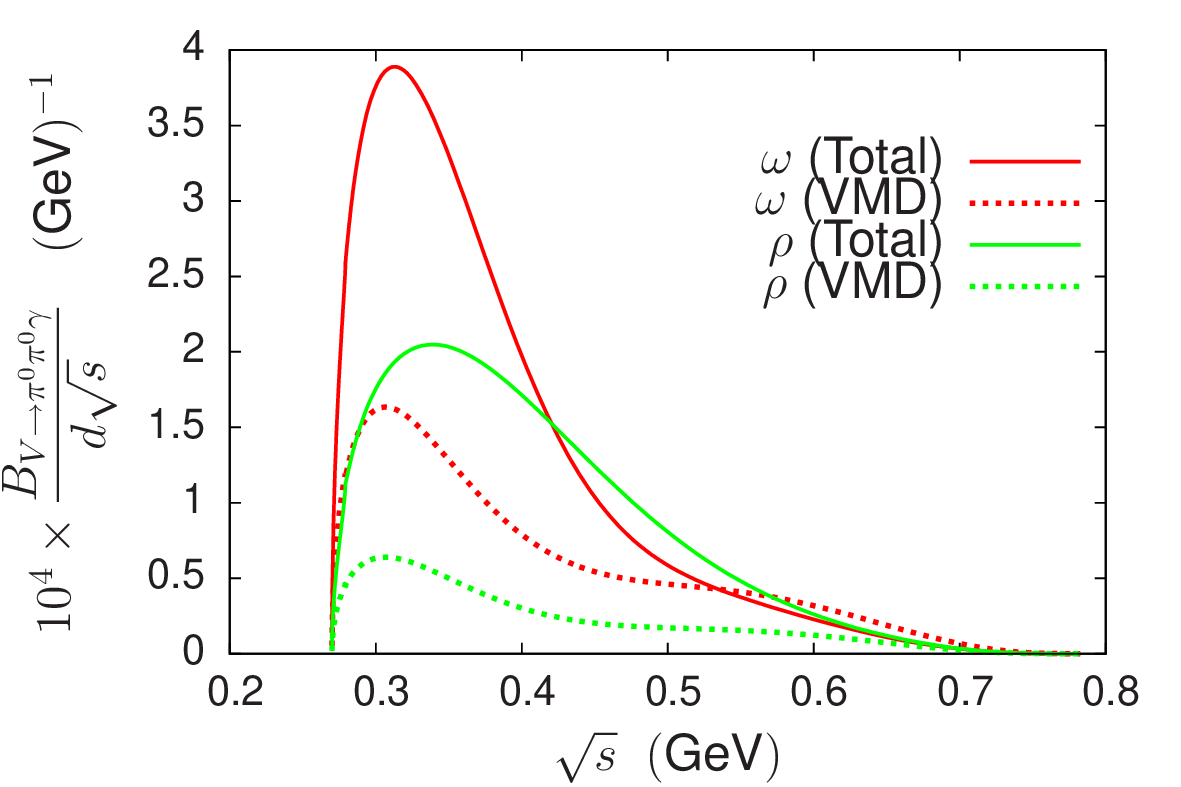}
\includegraphics[width=0.7\linewidth]{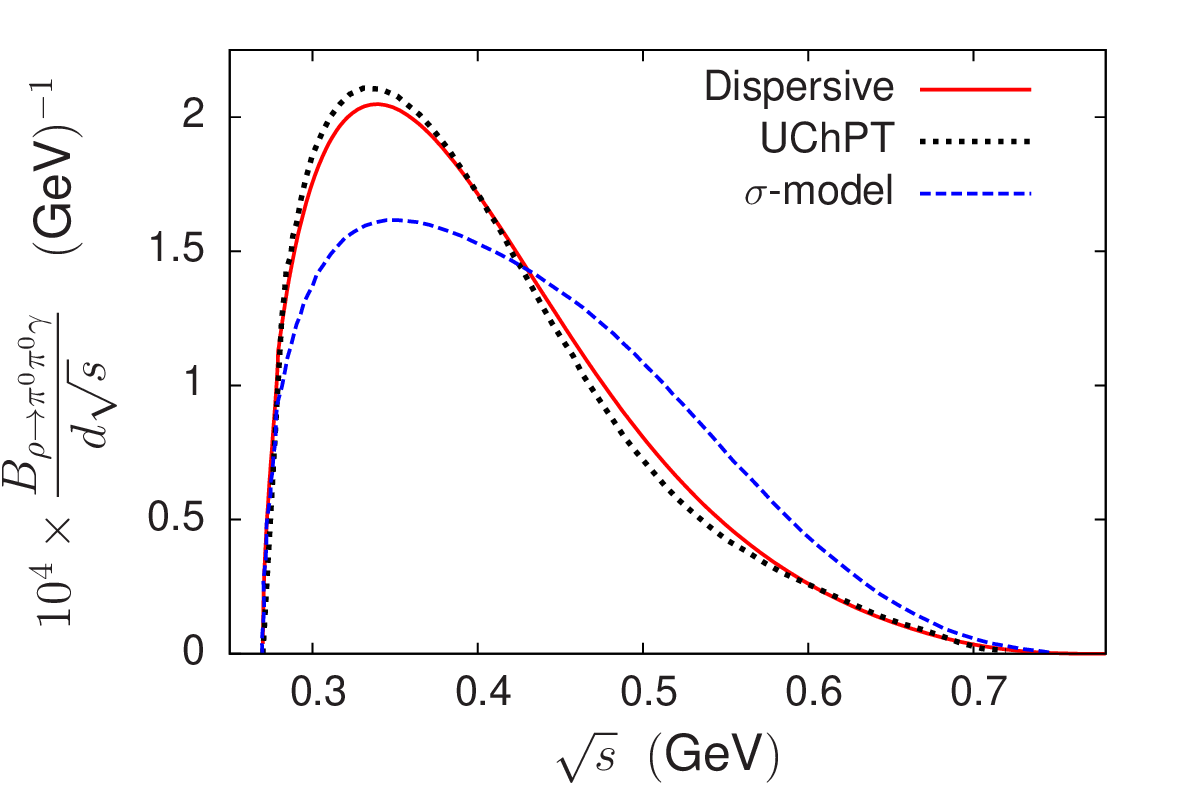}
\caption{\sl Upper plot: differential distributions, as a function of
  the $\pi\pi$ energy, of the branching fractions for
  $\omega\to\pi^0\pi^0\gamma$ and $\rho^0\to\pi^0\pi^0\gamma$ (solid
  lines). The dotted lines correspond to the amplitudes from the
  vector meson exchange diagrams alone. Lower plot: comparison of the
  results from the dispersive amplitudes and from other approaches: 
  ref.~\cite{Palomar:2001vg} (unitarized ChPT with resonances)
  and ref.~\cite{Escribano:2006mb} (sigma model). }  
\label{fig:diffVdecay}
\ec
\end{figure}

The shapes of the differential decay widths $d\Gamma_V/d\sqrt{s}$ of the
$\rho$ and the $\omega$ mesons, as a function of the $\pi\pi$ energy,
is illustrated on fig.~\fig{diffVdecay}. The two shapes are rather
different. This can be easily understood from the general structure of
the dispersive amplitudes~\rf{completehelic}. In the $\rho$ decay
amplitude, a large $\pi\pi$ rescattering contribution is induced from
the Born terms integrals $J^{I,\pi}$, which is absent in the isospin
limit for the $\omega$ decay amplitude. Our result for 
$d\Gamma_\rho/d\sqrt{s}$ is in better agreement with the
one obtained using a unitarized ChPT approach~\cite{Palomar:2001vg}
than those obtained using resonance models with a
Breit-Wigner $\sigma$-meson (see fig.~\fig{diffVdecay}).

Finally, let us quote our results for the decays into
$\pi^+\pi^-\gamma$. In this case, the Born amplitude contributes and
one must take the infrared divergence into account. We follow
ref.~\cite{Dolinsky:1991vq} and consider  the radiative 
width defined with a cutoff on the photon energy $E_\gamma \ge
E^{cut}_\gamma=50$ MeV. Integrating the differential decay
width~\rf{diffwidth} up to $s^{cut}=m_V^2-2m_VE^{cut}_\gamma$ we
obtain 
%updated 11 april
\be
\ba{lll}
BF(\rho\to\pi^+\pi^-\gamma)  =10.2\cdot10^{-3} & (\hbox{Born}), &
10.4\cdot10^{-3}\ (\hbox{Full})\\
BF(\omega\to\pi^+\pi^-\gamma)=1.85\cdot10^{-4}& (\hbox{Born}),&
2.59\cdot10^{-4}\ (\hbox{Full})\\
\ea
\en
The Born amplitude  dominates this mode in the case of the $\rho$
decay.  In the case of $\omega$ decay, the Born amplitude is
suppressed by isospin but its relative contribution is nevertheless
sizable. The experimental values for these branching fractions
are~\cite{Beringer:1900zz}
\be
\left.BF(\rho\to\pi^+\pi^-\gamma)\right\vert_{exp}=(9.9\pm1.6)\,10^{-3},\
\left.BF(\omega\to\pi^+\pi^-\gamma)\right\vert_{exp}< 36\cdot
10^{-4}\ .
\en

%%%%%%%%%%%%%figure vectors differential decay width

\subsection{Generalized polarizabilities}
\begin{figure}[ht]
\bc
\includegraphics[width=0.8\linewidth]{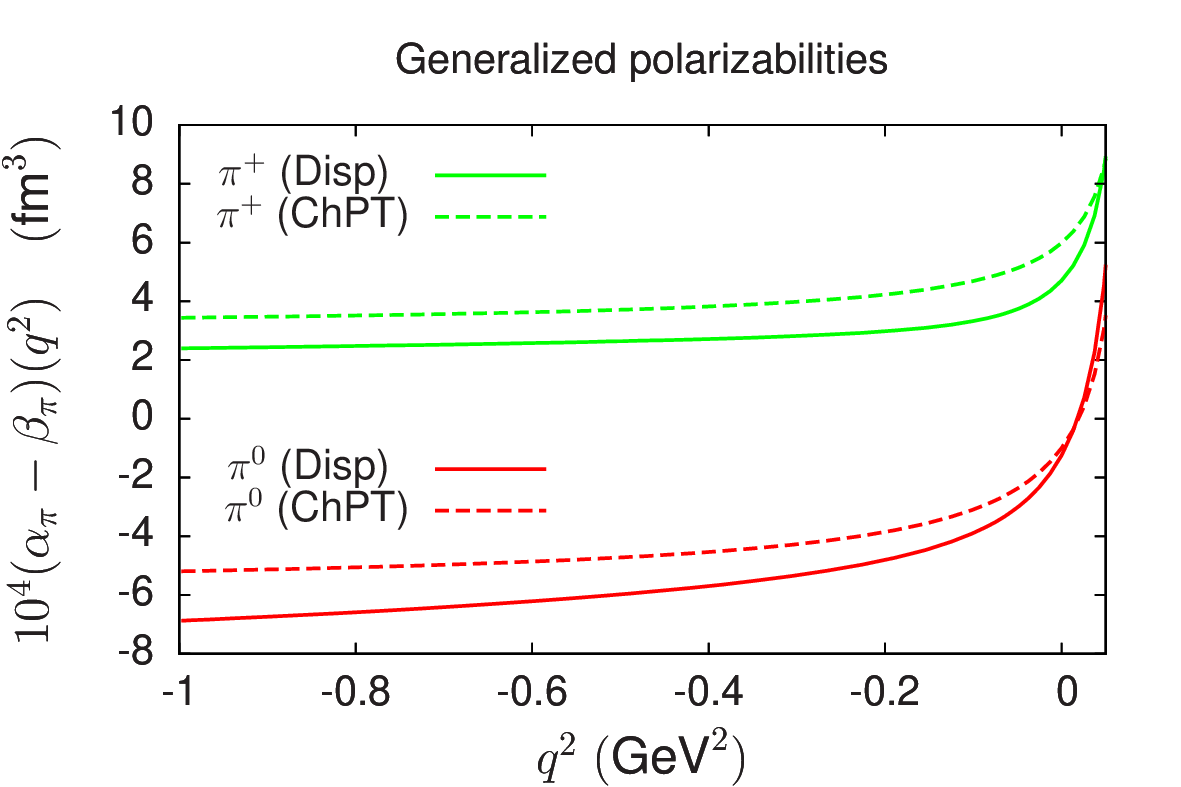}
\caption{\sl Generalized polarizability difference from the dispersive
amplitudes compared with chiral NLO result.}
\label{fig:polardiff}
\ec
\end{figure}

In the case where $\qdd < 0$, it
is fruitful to introduce the notion of generalized polarizabilities. This was
originally proposed in the case of the nucleon in
ref.~\cite{Guichon:1995pu} and  extended to the case of the pion and
further discussed in refs.~\cite{Unkmeir:1999md,L'vov:2001fz}. As
explained in ref.~\cite{L'vov:2001fz}, the generalized polarizabilities
characterize the spatial distribution in the hadron of the
polarizability induced by an external static electric or magnetic field.
These observables can be related to the coefficient
functions $A(s,t,\qdd)$, $B(s,t,\qdd)$,  $C(s,t,\qdd)$ in the limit
$t\to\mpid$, $s\to\qdd$ after subtracting the Born term.  We will be
concerned here with the polarizability difference, which is given
by~\cite{Unkmeir:1999md}  
\be
\alpha_\pi(\qdd)-\beta_\pi(\qdd)=\lim_{s\to\qdd,\,t\to\mpid}
{\alpha\over\mpi}\left(
A(s,t,\qdd)-2(s-4\mpid)B(s,t,\qdd) \right)
\en
($\alpha_\pi(\qdd)$ is denoted as $\alpha^L_\pi(\qdd)$ in
ref.~\cite{Unkmeir:1999md}). 
Considering the expression for the helicity amplitude
$H_{++}$ in terms of the coefficient functions~\rf{helicrels} 
one sees that the polarizability difference can be related to the
helicity amplitude taken in the limit $s\to\qdd$, $\theta\to\pi/2$
\be
\alpha_\pi(\qdd)-\beta_\pi(\qdd)=\lim_{s\to\qdd,\,\theta\to\pi/2}
{2\alpha\over\mpi} {\hat{H}_{++}(s,\qdd,\theta)\over s-\qdd}\ .
\en
The results deriving from our dispersive amplitudes are shown in
fig.~\fig{polardiff} and compared to the chiral NLO results. Those
have the simple expressions~\cite{Unkmeir:1999md}
\bea\lbl{polardiffNLO}
&& \left. \alpha_\piz(\qdd)-\beta_\piz(\qdd)\right\vert_{NLO}
={2\alpha\over\mpi} {(\qdd-\mpid)\over \fpid} \bar{J}'(\qdd)
\nonumber\\
&&  \left. \alpha_\pip(\qdd)-\beta_\pip(\qdd)\right\vert_{NLO}=
{2\alpha\over\mpi}\left( 
{\qdd\over 2\fpid} \bar{J}'(\qdd)+ {\bar{l}_6-\bar{l}_5\over
      48\pi^2\fpid}\right)\ . 
\ena
Keeping in mind  that the values at $\qdd=0$ in the dispersive
amplitudes have been chosen to be slightly different from the chiral
NLO values, fig.~\fig{polardiff} shows that the variation as a
function of $\qdd$ of the generalized polarizabilities is described by
the simple NLO expressions~\rf{polardiffNLO} to a rather good approximation.
%%%%%%%%%%%%%figure: generalized polarizabilities.
%%%%%%%%%%%%%figure: generalized polarizabilities.

\subsection{Sigma meson electromagnetic form factor}
The $\sigma$ meson resonance is often used as a simplified description
of the dynamics of $\pi\pi$ re-scattering in the isoscalar $S$-wave.
From this point of view, electromagnetic properties of this resonance
play a role in the hadronic contributions to the muon $g-2$. For
instance, in ref.~\cite{Narison:2003ur} a contribution to the vacuum
polarization was estimated assuming a vector dominance behaviour for
the $\gamma-\sigma$ form factor. This form factor would also be involved if
one considered the pole contribution
$\gamma\gamma^*\to\sigma\to\gamma^*\gamma^*$ in the light-by-light
scattering amplitude. 
In the approach used here, the $\pi\pi$ rescattering dynamics is expressed
in terms of the partial-wave $S$-matrix. The
sigma meson can be identified as a pole of this function, in the complex
energy plane, on the second Riemann sheet. It was shown in
ref.~\cite{Caprini:2005zr} that a rather precise determination can be
achieved, based on the Roy equations, despite the fact that this
resonance has a rather large width. Further work on this topic  was
done in ref.~\cite{GarciaMartin:2011jx}. 

%
%%%%%%%%%%%%%%%figure sigma-gamma form factor
\begin{figure}[hb]
\bc
\includegraphics[width=0.50\linewidth]{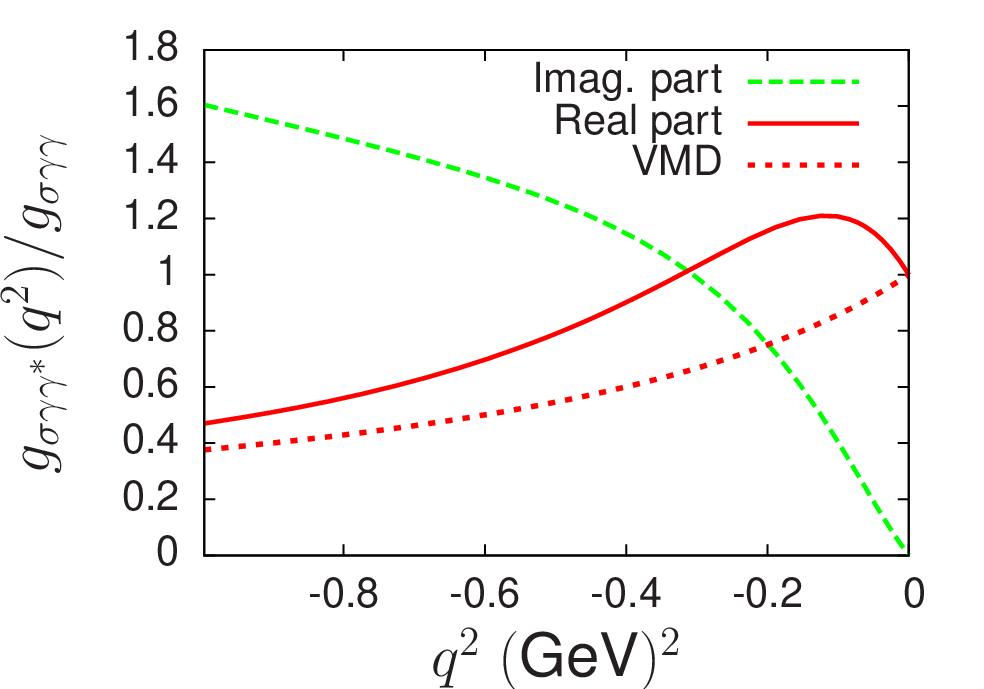}
\includegraphics[width=0.50\linewidth]{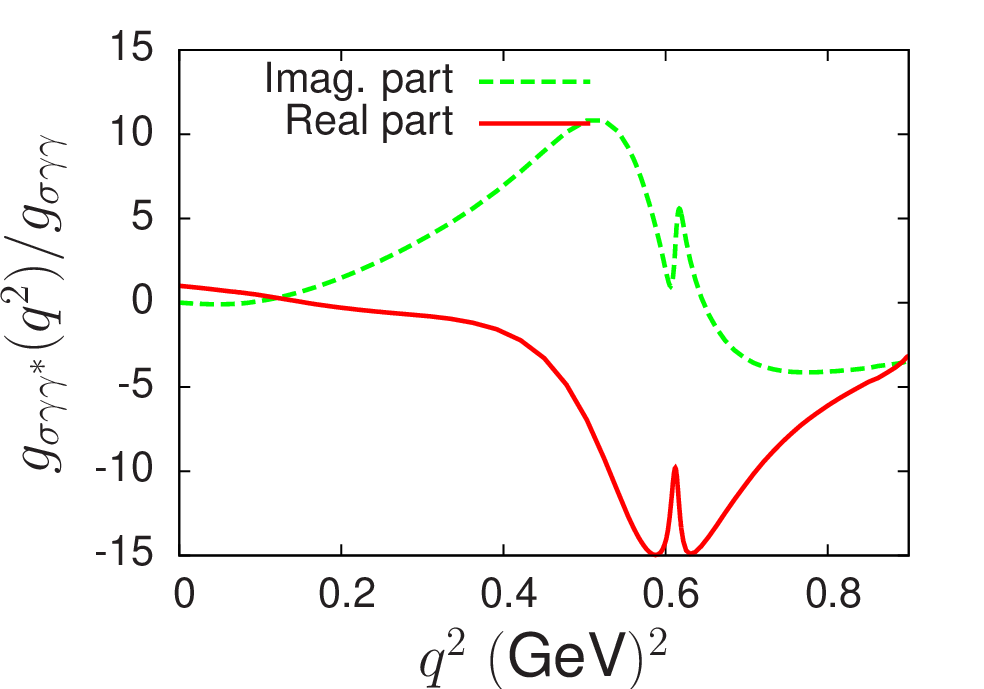}
\caption{\sl Real and imaginary parts of the ratio
  $g_{\sigma\gamma\gamma^*}(\qdd)/g_{\sigma\gamma\gamma^*}(0)$ as a
  function of the virtuality $\qdd$. }
\label{fig:gsigmaggstar}
\ec
\end{figure}
%%%%%%%%%%%%%%%figure sigma-gamma form factor
The discontinuity/unitarity relation shows that the second sheet poles
of the $\pi\pi$ $S$-matrix are also present in production amplitudes
such as $\gamma\gamma\to\pi\pi$, $\gamma\gamma^*\to\pi\pi$. 
The determination of the $q^2$ dependent form factor
$g_{\sigma\gamma\gamma^*}$ from the latter amplitude can be  performed in
exactly the same way as that of  the coupling constant
$g_{\sigma\gamma\gamma}$ from the former amplitude~\cite{Pennington:2006dg}
using the residues of the second-sheet poles,
\be\lbl{residpoles}
 \left.{\cal T}^{I=0}_{\pi\pi\to\pi\pi}\right\vert_{pole}= 
{\left(g_{\sigma\pi\pi}\right)^2 \over s_\sigma- s}\ ,
\quad
 \left. h^{I=0}_{++}(s,\qdd)\right\vert_{pole}=
 {g_{\sigma\gamma\gamma^*} g_{\sigma\pi\pi}\over s_\sigma- s}\ .
\en
This definition would correspond to the following matrix element 
of the electromagnetic current in the zero width limit for the $\sigma$,
\be
\braque{\gamma(q_1)\vert j_\mu(0)\vert \sigma(l_1)}= (\epsilon_1\cdot l_1
q_{1\mu} -q_1\cdot l_1\epsilon_{1\mu}) g_{\sigma\gamma\gamma^*}( (q_1-l_1)^2)\ .
\en
Using eq.~\rf{residpoles}, together with the fact that $s_\sigma$
corresponds to a zero of the $S$-matrix on the first sheet, on obtains
\be\lbl{gsigggformule}
g_{\sigma\gamma\gamma^*}(\qdd)= h^{I=0}_{++}(s_\sigma,\qdd)
\left( {-\tilde{\sigma}(s_\sigma)\over16\pi\dot{S}_0^0(s_\sigma)}
\right)^{1\over2}\ ,
\en
with $\tilde{\sigma}(z)=\sqrt{4\mpid/z-1}$. 
We use here an $I=J=0$ $\pi\pi$ $S$-matrix constrained by the Roy
equation up to the $K\bar{K}$ threshold discussed in~\cite{Moussallam:2011zg},
which gives (central values),
\be
s_\sigma=0.1202+i\,0.2422\ \hbox{GeV}^2\quad
\dot{S}_0^0(s_\sigma)=0.7573+i\,2.2055 \ \hbox{GeV}^{-2}\ .
\en
At $q^2=0$, firstly, we obtain from the
amplitude~\rf{completehelic},~\rf{biq2dep} the $\sigma$ to two photons
coupling
\be
g_{\sigma\gamma\gamma}=-3.45+i\,5.90\ \hbox{MeV}
\en
in reasonable agreement with our previous result~\cite{Moussallam:2011zg}
($g_{\sigma\gamma\gamma}=-3.14+i\,6.03$ MeV) based on a coupled channel
Omn\`es representation and a more complete description of the
left-hand cut. The variation as a function of  $\qdd$ is illustrated
on fig.~\fig{gsigmaggstar} which shows the ratio
$g_{\sigma\gamma\gamma^*}(\qdd)/g_{\sigma\gamma\gamma}$, separately
for $\qdd>0$ and $\qdd<0$. In a simple vector dominance  picture,
this ratio is expected to be proportional to $m_\rho^2/(m_\rho^2-\qdd)$ for
negative $\qdd$ implying that
$g_{\sigma\gamma\gamma^*}(\qdd)/g_{\sigma\gamma\gamma}$ should be
real in this region.  Fig.~\fig{gsigmaggstar} shows that the actual
results do not follow this simple VMD picture.

\subsection{Contribution to the muon anomalous magnetic moment}
Let us consider, finally, the  contribution from the cross-section
$\sigma(e^+e^-\to\gamma^*\to\gamma\pi^+\pi^-,\gamma\pi^0\pi^0)$  to
the anomalous magnetic  moment of the muon, $a_\mu=(g-2)/2$.  
This contribution was discussed, in the case of charged pions, in
refs.~\cite{De Troconiz:2001wt,Melnikov:2001uw} in the so called sQED
approximation, which corresponds to  retaining only the Born terms in
the expression of $\gamma^*\to\gamma\pi^+\pi^-$ amplitude.   
In the context of $a_\mu$, the range of validity of our 
amplitude $e^+e^-\to\gamma\pi\pi$ allows us to evaluate the
corresponding contribution in the range $q \le q_{max}=0.95$ GeV. Generically,
these contributions to $a_\mu$ have the following form
\be\lbl{hvpintegral}
 a_\mu^{[\gamma\pi\pi]}={1\over 4\pi^3}
\int_{4\mpid}^{q^2_{max}} d\qdd\, K_\mu\left({\qdd}\right)\,
\sigma_{e^+e^-\to \gamma\pi\pi}(\qdd)
\en
where the kernel function is compactly expressed as~\cite{Lautrup:1974ic}
\be
K_\mu(z)=\int_0^1 dx {x^2(1-x)\over x^2+  {z\over m^2_\mu}(1-x)}
\en
(see e.g.~\cite{Jegerlehner:2009ry} for the analytic expression and a
detailed review). 

In the case of charged pions, at first, one must take care of the
infrared divergence. This may be done by separating the contribution from
the Born amplitude squared 
\be
\vert H^c_{\lambda\lambda'}\vert^2= \vert
H^{Born}_{\lambda\lambda'}\vert^2
+2\re\left[{H_{\lambda\lambda'}^{Born}}^*\hat{H}^c_{\lambda\lambda'}\right]
+\vert\hat{H}^c_{\lambda\lambda'}\vert^2
\en
in the general expression for the cross-section~\rf{d2sigma} and
correspondingly writing the cross-section as a sum of three terms 
\be\lbl{separborn}
\sigma_{e^+e^-\to \gamma\pi\pi}(\qdd)=\sigma^{Born}(\qdd)
+\hat{\sigma}^{Born}(\qdd)+\hat{\sigma}(\qdd)\ .
\en
Only the first term in eq.~\rf{separborn} is affected by an infrared
divergence. For definiteness, let us consider the inclusive
definition, where $\sigma^{Born}(\qdd)$ is defined by combining it
with the radiative correction to the vertex
$\gamma^*\to\pi^+\pi^-$, and can then be written as follows
\be
\sigma^{Born}(\qdd)={\pi\alpha^2\over 3\,\qdd}\,\betapi^3(\qdd)\,\vert
F^v_\pi(\qdd)\vert^2 \times {\alpha\over\pi}\eta(\qdd)
\en
(see e.g. ref.~\cite{Jegerlehner:2009ry} where the explicit expression for
the function $\eta(\qdd)$ can be found). 
In accordance with the  decomposition~\rf{separborn} of the
cross-section we can write the contributions from $\gamma\pi^+\pi^-$
to the muon anomalous magnetic moment as 
\be
a_\mu^{[\gamma\pi^+\pi^-]}=
a_\mu^{Born}+\hat{a}^{[\gamma\pi^+\pi^-]}_\mu\ ,
\en 
and we find the following numerical results
\be\lbl{amucharged}
\ba{ll}
a_\mu^{Born}& \!\!\!=41.9\cdot10^{-11}\\
\hat{a}^{[\gamma\pi^+\pi^-]}_\mu & \!\!\!=(1.31+0.16\pm0.40)\cdot10^{-11}\ ,\\
\ea
\en
showing the separate contributions from the three terms
in eq.~\rf{separborn}. 
Similarly, we can compute the contribution to the muon anomalous
magnetic moment from the neutral channel $\gamma\pi^0\pi^0$, and we find
\be\lbl{amuneutral}
a_\mu^{[\gamma\pi^0\pi^0]}=(0.33\pm 0.05)\cdot10^{-11}\ .
\en
As was noted in the literature~\cite{De
  Troconiz:2001wt,Melnikov:2001uw,Czyz:2003ue}) the contribution of
the $\gamma\pi^+\pi^-$ channel from the purely Born 
terms, $a_\mu^{Born}$  is not negligible. It is of the same size as the 
present error in the Standard Model evaluation~\cite{Davier:2010nc},
$\Delta a_\mu=\pm4.9\cdot10^{-10}$.  Comparatively, the other
contributions from $\gamma\pi\pi$ are rather small, even though they
do include some enhancement from the strong $\pi\pi$ rescattering
in the isoscalar $S$-wave. As an effect of rescattering, the
contribution from the term linear in the Born amplitude (second term
in eq.~\rf{separborn}) is found to be positive here, contrary to the result
of~\cite{Dubinsky:2004xv}. 
Our evaluation of $\hat{a}^{[\gamma\pi\pi]}_\mu$~\rf{amucharged},~\rf{amuneutral} 
should be more precise than the estimates using $\sigma$-meson
approximations~\cite{Narison:2003ur,Ahmadov:2010hq}. 

\section{Conclusions}
We have discussed the generalization of dispersive Omn\`es-type
representations of $\gamma\gamma\to\pi\pi$ amplitudes
(as e.g. in refs.~\cite{Morgan:1991zx,Donoghue:1993kw}) to the case
where one photon is virtual.  
These approaches involve a modelling of the left-hand cut, beyond the
pion pole contribution, in terms of light resonance exchanges. We
showed how this can be consistently defined as a generalized
left-hand cut, with no intersection with the unitarity cut, through
the use of a K\"allen-Lehmann representation for the resonance
propagators and the limiting $i\epsilon$ prescriptions for energy
variables. 
Our main result is a representation of $\gamma^*\to\gamma\pi\pi$ 
(or $\gamma^*\gamma\to\pi\pi$) helicity amplitudes,
eq.~\rf{completehelic} which is  based on  twice-subtracted
dispersion relations for the $J=0$ partial-waves.  The representation
satisfies the soft photon
theorem and displays explicitly the dependence on the $\pi\pi$ $I=0,2$
phase-shifts, on the pion electromagnetic form factor and on $V\pi$
form factors. It also involves two functions of the virtuality
$\qdd$. These functions are constrained  by matching with ChPT,
through the values at $\qdd=0$ and their relation to the pion
polarizabilites. We then showed that a simple two parameter
representation is adequate for reproducing the experimental data on
$e^+e^-\to\gamma\pi^0\pi^0$. Eq.~\rf{completehelic} is valid in a
range of virtualities $\qdd$ and $\pi\pi$ energies not exceeding 1
GeV, such that $\pi\pi$ scattering is essentially elastic and the
phase-shifts of $J\ge 2$ partial-waves may be neglected. In principle,
it is possible to extend this kind of representation to somewhat
higher energies where inelasticity is dominated by the single
$K\bar{K}$ channel, by constructing numerical solutions to the
coupled Muskhelishvili-Omn\`es equations. 

As a first application, the behaviour of the
generalized polarizability difference $\alpha_\pi-\beta_\pi$, as a
function of $q^2$, was derived. This function is found not to deviate
much from the prediction of ChPT at NLO at negative $\qdd$.
As a second application, results were deduced  for vector meson decay
amplitudes $\rho,\omega\to\gamma\pi\pi$ 
as well as the $\gamma\sigma$ electromagnetic form
factor. This latter object can be defined from the second sheet pole
definition of the $\sigma$ resonance and generalizes the
$\sigma\gamma\gamma$ coupling introduced in
ref.~\cite{Pennington:2006dg}. 
The amplitudes $\gamma^*\to\gamma\pi\pi$ participate in the hadronic
vacuum polarization contribution to the muon $g-2$. We have evaluated
the contributions beyond the point-like approximation, which is usually
accounted for. 

\section*{Acknowledgements}
I would like to thank Diogo Boito for his participation in the first
stages of this project.
Work supported in part by the European Community-Research
Infrastructure Integrating Activity "Study of Strongly Integrating
Matter" (acronym HadronPhysics3, Grant Agreement n. 283286) under the
Seventh Framework Programme of EU. 

%%%%%%%%%%%%%%
\section*{Appendices}
\appendix
%\appendixpage
%\addappheadtotoc

%%%%%%%%%%%%%%figure:comp BWtilde and GS
\begin{figure}[hbt]
\bc
\includegraphics[width=0.8\linewidth]{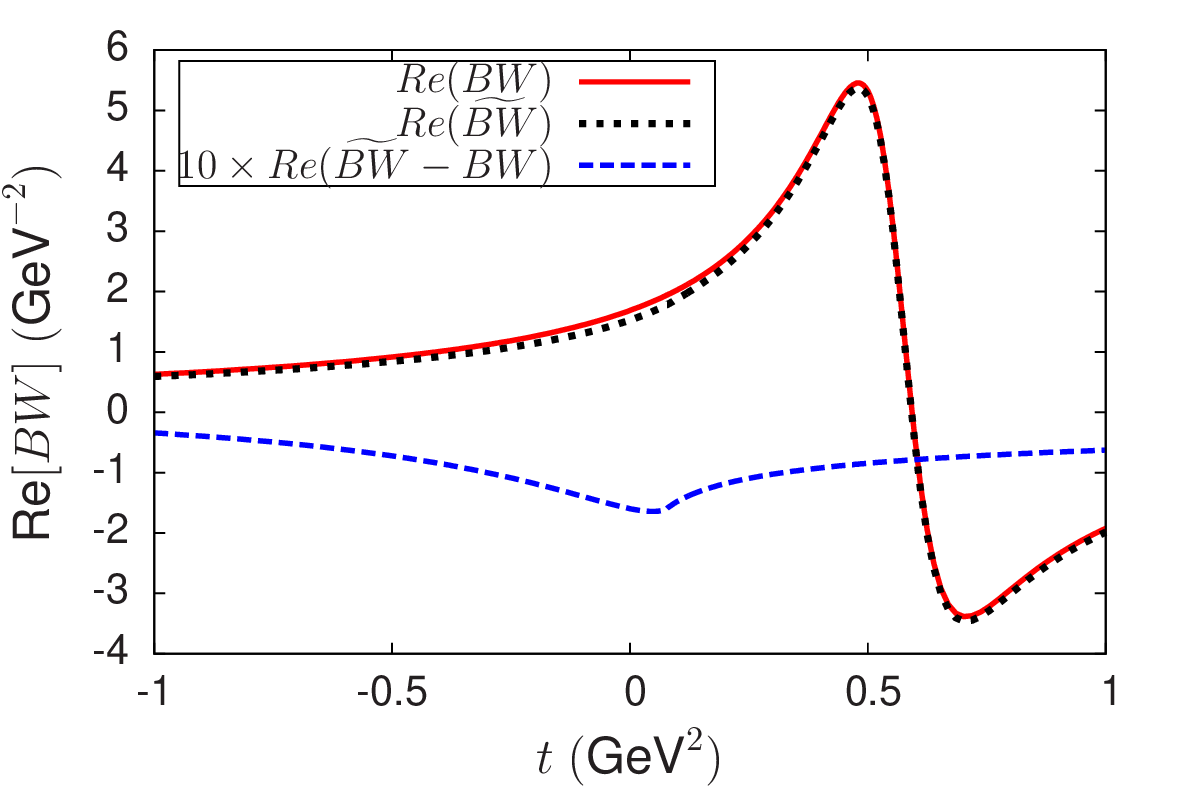}
\caption{Comparison of the real part of the usual 
  $\rho$-meson propagator $BW(t)$ and the real part of modified propagator
  $\widetilde{BW}(t)$ which has correct analyticity properties.} 
\label{fig:kstilde}
\ec
\end{figure}
%%%%%%%%%%%%%%figure:comp BWtilde and GS
\section{Vector resonance propagator with good analyticity
  properties}\lblsec{Vpropagator} 
We give here an explicit representation for the resonance propagator
$\widetilde{BW}(t)$ given from eqs.~\rf{spectralprop} and
~\rf{kstilde} (see also~\cite{Lomon:2012pn}). 
In order to evaluate the integral ~\rf{spectralprop} in
analytic form, one must first compute the three zeros $t_R$, $t_\pm$
of the denominator of $\im[BW(t)]$. Approximate values are,
\be
t_R\simeq \mvd {\epsilon_V^3 \gamma_V^2 \over 
1 + 3\epsilon_V^2 \gamma_V^2},\ 
t_\pm\simeq {m^2_V\over 1 \pm i\gamma_V}\ 
\en
with $\epsilon_V=4\mpid/m^2_V$.
More precise values must be determined numerically. 
The spectral integral can then be expressed in terms of the loop function
$\bar{J}_\pi(z)$ (see~\rf{Jbar}), 
as
\be
\widetilde{BW}(t)={16\pi\gamma_V\over 1+\gamma^2_V}\left[
 A(t)\, \bar{J}_\pi(t)+ B(t)\, \bar{J}_\pi(t_R) 
+ C_+(t)\bar{J}_\pi(t_+)+ C_-(t)\bar{J}_\pi(t_-)  \right]\ ,
\en
where the coefficient functions $A$, $B$, $C_\pm$ are given by
\be
\begin{array}{l}
A(t)={\displaystyle t(t-4\mpid)\over\displaystyle(t-t_R)(t-t_+)(t-t_-)}\\
B(t)={\displaystyle t_R(t_R-4\mpid)\over\displaystyle (t_R-t)(t_R-t_+)(t_R-t_-)}\\
C_\pm(t)={\displaystyle t_\pm(t_\pm-4\mpid)\over\displaystyle (t_\pm-t)(t_\pm-t_R)(t_\pm-t_\mp)}\\
\end{array}
\en
and satisfy $A+B+C_++C_-=0$. The poles in these functions 
cancel in $\widetilde{BW}(t)$, but they are present on the second
Riemann sheet, which is easily seen using the second sheet extension
of $\bar{J}$,
\be
\bar{J}^{II}(t)= \bar{J}(t)+{2i\sigma_\pi(t)\over16\pi}\ .
\en
Fig.~\fig{kstilde} compares the real
parts of  $BW(t)$ and  $\widetilde{BW}(t)$.

\section{Probing the dispersive formulae with simple triangle diagrams}
\begin{figure}[th]
\bc
\includegraphics[width=6cm]{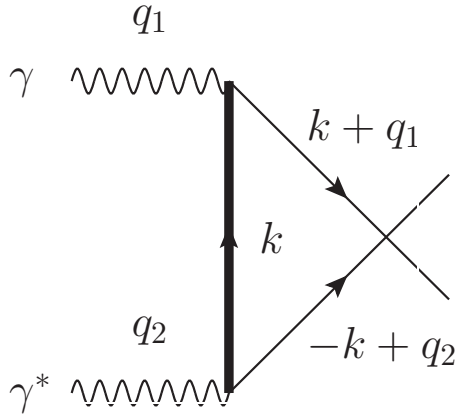}
\caption{Simple triangle diagram.}
\label{fig:triangle}
\ec
\end{figure}

\lblsec{Disptriangles}
We   consider  here simple triangle diagrams (fig.~\fig{triangle}) as
a toy model of rescattering.  This allows one to check the absence
of anomalous thresholds in the dispersive representation and the
correctness of the prescriptions for calculating the real and imaginary
parts of the amplitude. All   particles in the diagrams
are spinless  but we  take kinematical   conditions   analogous  to
those   relevant  for   our $\gamma\gamma^*$ problem. We take  the
mass of the particle associated with the vertical  line to be $M$  and
the other two masses  to be $m$. The amplitude can be expressed as a
one-dimensional parametric integral, 
\be\lbl{triparam}
 {\cal T}={-1\over16\pi^2}
 \int_{0}^1 
{d\alpha\over \alpha(s-\qdd)+ M^2 -m^2}
\log{m^2-\alpha(1-\alpha)s\over \alpha
  m^2+(1-\alpha)M^2-\alpha(1-\alpha)\qdd}  \ .
\en
It is easily verified that the denominator produces no singularity,
the only singularities are contained in the logarithms.
The real and imaginary parts of the amplitude, in the
representation~\rf{triparam}  correspond to  integrating over  the real
part and the imaginary part of  the logarithm. In order to define the
proper sign for the imaginary parts the energy variable $s$ is
considered as the limit of $s+i\epsilon$ and similarly for the energy
variable $\qdd$.
%
%%%%%%%%%%%%%%%figure Feynman triangle
%%%%%%%%%%%%%%%figure Feynman triangle
Let us now consider two cases for the masses\\[0.5cm]
{\bf a) $M=m$: }\\
In this situation, the parametric representation simplifies to
\be\lbl{eqmparam}
{\cal T}={-1\over16\pi^2(s-\qdd)}\int_{0}^1 {d\alpha\over\alpha}
\log{m^2-\alpha(1-\alpha)s\over m^2-\alpha(1-\alpha)\qdd}\ .
\en
Let us examine the dispersive  representation in the
variable $s$. The discontinuity is easily found to be
\be
\disc_s{\cal T}\equiv { {\cal T}(s+i\epsilon)-{\cal
    T}(s-i\epsilon)\over 2i}
= {\theta(s-4m^2)\over 16\pi (s-\qdd)} \log{ 1+\sigma(s)\over 1-\sigma(s)}
\en
with $\sigma(s)=\sqrt{1-4m^2/s}$.
The discontinuity $\disc_s{\cal T}$ has a structure similar to the QED
Born term~\rf{pwborn0} (in particular, it has a singularity at $s=\qdd$). 
The dispersive representation of the amplitude  requires
no subtraction and has the form
\be\lbl{eqmdisp}
{\cal T}(s)= {1\over\pi}\int_{4m^2}^\infty {ds'\over s'-s}
\disc_s{\cal T}(s')\ .
\en
One can verify that the representation~\rf{eqmdisp} is correct
(i.e. the absence of an anomalous threshold), as
it can be derived from~\rf{eqmparam} by making a simple change of
variable. Splitting the integration range into two parts:
$[0,{1\over2}]$ and  $[{1\over2},1]$, one sets $\alpha=\alpha_-(s')$ in the
first range and $\alpha=\alpha_+(s')$ in the second, with 
\be
\alpha_\pm(s')={1\over2}\left( 1\pm \sigma(s')\right)\ .
\en

We can rewrite this representation in a form which exhibits the
symmetry in $s$ and $\qdd$, in terms of a difference of two integrals 
\be
{\cal T}= {I(s) -I(\qdd)\over s-\qdd}
\en
with
\be
I(x)={x\over 16\pi^2}\int_{4m^2}^\infty {ds'\over s'(s'-x)}
\log{1+\sigma(s')\over 1-\sigma(s')}\ .
\en
This form is a simplified analog of the Born term rescattering piece
in eq.~\rf{OMformule}. The function $I(x)$ can be expressed analytically
%real part sign corrected 22 juill.
\be
I(x)=\left\{
\ba{ll}
x\le 0 & -{1\over32\pi^2}\log^2 {\beta(x)+1\over \beta(x)-1}\\
0\le x\le 4m^2 & {1\over 8\pi^2} \arctan^2\sqrt{x\over 4m^2-x}\\
x\ge 4m^2 & -{1\over32\pi^2}\left( \log{1+\beta(x)\over1-\beta(x)}-i\pi
\right)^2\ .
\ea \right. 
\en 
Let us finally remark that the imaginary
part  of the  amplitude  ${\cal  T}$  does not  necessarily coincide
with  the  $s$-discontinuity,  depending  on  the  value  of
$\qdd$. Indeed,  
\be
\ba{ll}
\im{\cal T}=\dfrac{\im I(s)}{s-\qdd}=\disc{\cal T}\ ,  
& \qdd \le 4m^2\\[0.3cm]
\im{\cal T}=\dfrac{\im I(s)-\im I(\qdd)}{s-\qdd}\ne\disc{\cal T}\ ,  
& \qdd >  4m^2\ .
\ea
\en

\noindent{\bf b) $M\ne m$: }\\
Let  us  now  consider  the  unequal  mass  case.  Starting  from  the
parametric representation~\rf{triparam} one obtains
the expression for the $s$-discontinuity as,
\be
\disc_s{\cal T}= {\theta(s-4m^2)\over16\pi (s-\qdd)} L_V(s,\qdd,M^2)
\en
where the function $L_V$ is the same which appears in the 
vector-exchange amplitude~\rf{Lv}. Thus, the dispersion relation
representation of ${\cal T}$ reads,
\be\lbl{disptriangle}
{\cal T}= {1\over 16\pi^2}
\int_{4m^2}^\infty {ds'\over (s'-s)(s'-\qdd)}\,L_V(s',\qdd,M^2)\ .
\en
We have not been able to derive this expression by making a simple change
of variables as in the equal mass case. As a check of its correctness,
we show below that the imaginary parts of the dispersive and
parametric representations coincide. For the real parts, we
checked their equality only numerically.

When using eq.~\rf{disptriangle} one  must be careful in the
evaluation of the real and  imaginary parts  of  the
amplitude. Unlike the case when $M=m$, no imaginary part is generated
from the  $s'-\qdd$ denominator (because $L_V(\qdd,\qdd)=0$). 
The function $L_V(s',\qdd,M^2)$, on the other hand,  has an imaginary
part if $\qdd > (M+m)^2$. With $s'$ real and $\qdd$ the limit of
$q^2+i\epsilon$,  the proper definition of $L_V$ is  
\be
L_V(s',\qdd,M^2)=\re L_V(s',\qdd,M^2) -i\pi\theta((s'-s_-)(s_+-s')) \ .
\en
The explicit expressions for $s_\pm(\qdd,M^2)$ were given in
eq.~\rf{splusmoin1}. Inserting this in the dispersive representation of
the triangle amplitude~\rf{disptriangle} we get, for the imaginary part 
\bea\lbl{imTdisp}
&& \im{\cal T}(s,\qdd)_{disp}= {1\over 16\pi}\bigg( 
\theta(s-4m^2)\re L_V(s,\qdd,M^2) \nonumber\\
&&\quad -\theta(\qdd-(M+m)^2){1\over
  s-\qdd}\log\left\vert { (s_+-s)(s_- -\qdd)
\over (s_--s)(s_+-\qdd)}\right\vert
\bigg)\ .
\ena 
Let us verify that this result coincides with the one obtained from
the parametric representation. We can write the logarithm in the
integrand in  eq.~\rf{triparam} as a difference: 
$\log({\cal P}_s(\alpha))-\log({\cal Q}_{\qdd}(\alpha))$. The zeros of
the polynomial ${\cal P}_s(\alpha))$ are given by  
\be
\alpha_\pm(s)= {1\over2}\left(1\pm \sigma(s) \right)
\en
and $\log({\cal P}_s(\alpha))$ has an imaginary part when $s>4m^2$ and
$\alpha_-(s) \le \alpha\le \alpha_+(s)$. The zeros of the polynomial ${\cal
  Q}_{\qdd}(\alpha)$ are given by 
\be
\beta_\pm(\qdd)={1\over2\qdd}\left( \qdd + M^2- m^2 \pm \sqrt{
  \lambda(\qdd,M^2,m^2)} \right)
\en
and $\log({\cal Q}_{\qdd}(\alpha))$ has an imaginary part when $\qdd >
(M+m)^2$ and $\beta_-(\qdd) \le \alpha\le \beta_+(\qdd)$. Performing the
integration over $\alpha$ we find that eq.~\rf{triparam} gives
\bea\lbl{imTparam}
&& \im{\cal T}(s,\qdd)_{param}= {1\over16\pi} {1\over
  s-\qdd}\bigg(  \nonumber\\
&&\qquad\theta(s-4m^2)        
\log\left\vert{\alpha_+(s)(s-\qdd)+ M^2-m^2\over 
\alpha_-(s)(s-\qdd)+ M^2-m^2}\right\vert\nonumber\\
&&\qquad +\theta(\qdd-(M+m)^2)
\log\left\vert{\beta_-(\qdd)(s-\qdd)+ M^2-m^2\over 
\beta_+(\qdd)(s-\qdd)+ M^2-m^2}\right\vert \bigg)\ .
\ena
We can now compare eq.~\rf{imTparam} with eq.~\rf{imTdisp}: the two
expressions are  seen to be  identical upon using the
relations between the $s^\pm$ and the $\beta^\pm$ functions
\be
s_\pm(\qdd)-\qdd= -{ (M^2-m^2)\qdd\over M^2}\,\beta_\pm(\qdd),\quad 
\beta_+(\qdd)\beta_-(\qdd)={M^2\over \qdd}\ .
\en
The real part of the amplitude is more difficult to evaluate
analytically, but one can verify numerically that the real parts of
the dispersive and the parametric representations also coincide.
\section{Finiteness of $\hat{J}^\pi(\qdd)$ at the $K\Kbar$ threshold}\lblsec{Nodivergence}
We show here that no divergence affects the function
$\hat{J}^{\pi}(\qdd)$ (see~\rf{Jhat}) at the $K\Kbar$ threshold if
one uses a two channel formalism. In connection with two-channel
unitarity, one must use a $2\times2$ Omn\`es matrix $\mathbf{\Omega}$
and the formula for $\hat{J}^\pi(\qdd)$ becomes 
\be
\left(\ba{l}
\hat{J}^\pi(\qdd)\\
\hat{J}^K(\qdd)\\
\ea\right)
=-{1\over \pi}\int_{4\mpid}^\infty {ds'\over(s'-\qdd)}\,
{d\over ds'}\, \left(
{1\over(s')^2}\im \left( \mathbf{\Omega}^{-1} \right)
\left(\ba{l}
{\displaystyle4\mpid L_\pi(s')}-2\qdd\\
{\displaystyle4\mkd  L_K(s')}-2\qdd
\ea\right)\right)
\en
which replaces eq.~\rf{Jhat}. The inverse of the Omn\`es matrix
satisfies  the following unitarity relation
\be
\im \mathbf{\Omega}^{-1}=-\mathbf{\Omega}^{-1}\times\mathbf{T}\times
\left(\ba{cc}
\betapi(s')\theta(s'-4\mpid) & 0 \\
0 &    \betaK(s')\theta(s'-4\mkd)\\
\ea\right)
\en
where $\mathbf{T}$ is the $2\times2$ $T$-matrix. By construction,
multiplying the $T$-matrix by the inverse of the Omn\`es function
removes the right-hand cuts, so the matrix elements
$(\mathbf{\Omega}^{-1}\,\mathbf{T})_{ij}$ should not exhibit any cusp
at the $\pi\pi$ or the $K\Kbar$ thresholds and therefore have
continuous derivatives. The derivative of the remaining pieces involve
\be
{d\over ds'}\left[\sigma_P(s')\left(
4m_P^2 L_P(s')-2\qdd \right)\right]=2(s'-q^2)\,\dot\sigma_P(s')
\en
(with $P=\pi, K$) which give contributions which are finite and
independent of $\qdd$.

\section{Electromagnetic form factors}\lblsec{Formfactors}
\subsection{Pion form factor $F^v_\pi(\qdd)$}
We need a description of the modulus and phase of the pion
electromagnetic form factor $F^v_\pi(\qdd)$. The form factor is
defined from the matrix element in eq.~\rf{ffactordef}.
The  modulus  can  be determined from experiment in  the physical region 
($s \ge 4m_\pi^2$) since it is related to the $\e^+ e^-\to \pi^+
\pi^-$ cross-section by the formula 
\be\lbl{sigmaeepipi}
\sigma_{\e^+ e^-\to \pi^+ \pi^-}(\qdd)= \vert F^v_\pi(\qdd)\vert^2\,
{\pi\alpha^2 (\qdd+2\med)\,\sigma_\pi^3(\qdd)\over 3 (\qdd)^2
  \,\sigma_e(\qdd)} 
\en
at leading order in $e^2$.
Many such measurements have been performed recently, see
e.g~\cite{Lees:2012cj} and references therein.
Accurate  representations at medium energy can be obtained from a
simple superposition of Breit-Wigner-type
functions~\cite{Kuhn:1990ad}. We  use here the fit performed by the CMD-2
collaboration~\cite{akhmetshin2007}, based on the following
representation, 
\be\lbl{Fpireson}
F^v_\pi(\qdd)= {1\over 1+\beta}\left[ GS_{\rho(770)}(\qdd) \left(1 +
  \delta\, {\qdd\over m_\omega^2} BW_{\omega}(\qdd)\right) +\beta\,
  GS_{\rho(1450)}(\qdd) \right]\ .  
\en  
In eq.~\rf{Fpireson}, $GS_R$ is the Gounaris-Sakurai
function~\cite{GS}, which can be expressed as follows
\be
GS_R(\qdd)=\frac{D_R(0)}{D_R(\qdd)},\quad D_R(\qdd)=m_R^2-\qdd -\gamma_R\left(
F(\qdd)-F(m_R^2)-(\qdd-m_R^2)F'(m_R^2)\right) 
\en
with
\be
\gamma_R={m_R \Gamma_R\over \sigma_\pi(m_R^2)(m_R^2-4\mpid)},\ 
F(\qdd)=16\pi\,(\qdd-4\mpid)\bar{J}_\pi(\qdd)
\en
and the loop function $\bar{J}_\pi(\qdd)$ is given in eq.~\rf{Jbar}. For the
$\omega$ meson, a simple Breit-Wigner function is used in
eq.~\rf{Fpireson} 
\be
BW_\omega(s)={m_\omega^2\over m_\omega^2-s -im_\omega \Gamma_\omega}\ .
\en

\subsection{$F_{\omega\pi}(\qdd)$ form factor}
Naively, we expect that the $F_{\omega\pi}$ form factor should be
somewhat similar to the pion form factor, i.e. that it should be
approximated reasonably well by a superposition of $\rho(770)$ and
$\rho(1450)$ resonances with a small isospin violating contribution
from the $\omega$.
However, there could be  some differences for two reasons: 1) the
phase of $F_{\omega\pi}(\qdd)$ is not related to the $\pi\pi$
scattering phase $\delta_1^1(\qdd)$, unlike the phase of of
$F^v_\pi(\qdd)$ and  2) instability of the omega meson permits
triangle diagram contributions to the form factor which violate real
analyticity. In other terms, the discontinuity of the form factor
along the elastic cut is complex. A dispersion relation analysis of
$F_{\omega\pi}(\qdd)$ which  takes such effects into account 
was performed some time ago~\cite{Koepp:1974da}. This was reconsidered
more recently in ref.~\cite{Schneider:2012ez} whose dispersive
analysis is based on  self-consistent solutions of  Khuri-Treiman type
equations for the $\pi\pi\to\omega\pi$ scattering
amplitude~\cite{Niecknig:2012sj}. As compared to 
these results, the Breit-Wigner type approach appears to be,
at least qualitatively, acceptable and we will use it here because of
its simplicity. 
  
From an experimental point of view the form factor $F_{\omega\pi}(\qdd)$
has been probed in the region $q \ge m_\omega+m_\pi\simeq 0.92$
GeV from
$e^+e^-\to\omega\pi^0$~\cite{Dolinsky:1986kj,Bisello:1990du,Achasov:2000wy,Akhmetshin:2003ag,Achasov:2012zz}   
and from the $\tau$ decays $\tau^\pm\to\omega \pi^\pm \nu_\tau$ by the CLEO
collaboration~\cite{Edwards:1999fj}. It has also been measured in the
energy region  $q \le m_\omega-m_\pi\simeq 0.65$ GeV from
$\omega\to l^+ l^- \pi^0$
decays~\cite{Dzhelyadin:1980tj,Landsberg:1986fd,Akhmetshin:2005vy,Achasov:2008zz,Arnaldi:2009aa}. Let us first consider the
$e^+e^-$ and $\tau$ decay measurements. Experimental observables are
related to the form factor $F_{\omega\pi}(\qdd)$ by the following expressions
\bea
&&\sigma_{e^+ e^-\to \omega\pi^0}(\qdd)= \tilde{C}_\omega\vert
F_{\omega\pi}(\qdd)\vert^2\,\,{ 4\pi\alpha^2 
\lambda_{\omega\pi}^{\scriptstyle3/2}(\qdd)\over 3 (\qdd)^3 }
\\
&& {d\Gamma_{\tau^\pm \to \omega \pi^\pm \nu_\tau}(\qdd)\over d\qdd }=
\tilde{C}_\omega\vert F_{\omega\pi}(\qdd)\vert^2\,\,  
{V^2_{ud} G_F^2 m_\tau^3
  \lambda^{\scriptstyle3/2}_{\omega\pi}(\qdd)\over 96\pi^3 (\qdd)^2} 
\left(1-{\qdd\over\mtaud}\right)^2 
\left(1 +{2\qdd\over\mtaud}\right)\nonumber
\ena
with $\lambda_{ab}(\qdd)=(s-(m_a-m_b)^2)(s-(m_a+m_b)^2)$.
A  dimensionless quantity also shown in ref.~\cite{Edwards:1999fj} is
the $\omega\pi$ spectral function which is given by
\be
V_{\omega\pi}(\qdd)= \tilde{C}_\omega\vert F_{\omega\pi}(\qdd)\vert^2\,\,  
{\lambda_{\omega\pi}^{\scriptstyle3/2}(\qdd)\over 3\pi (\qdd)^2}\ .
\en
These experiments probe the ``tail'' of the $\rho(770)$ resonance and
can be reproduced in the region $\sqrt{s}\lapprox 1.5$ with a
resonance superposition model very similar to that of $F^v_\pi$
\be\lbl{Fomegapi0}
F_{\omega\pi}(\qdd)={1\over 1+\beta'} \left[  GS_{\rho(770)}(\qdd)
 \left(1+\delta {\qdd\over m_\omega^2} BW_{\omega}(\qdd)\right)
+\beta'  GS_{\rho(1450)}(\qdd) \right]\ .
\en
The parameter $\delta$ which describes $\omega-\rho$ mixing is taken
to be the same as in $F^v_{\pi}$. The parameter $\beta'$ is not
related to the corresponding one in $F^v_{\pi}$ because the phases
of the form factors $F^v_{\pi}$ and $F_{\omega\pi}$ should be
allowed to be  different.
We perform a fit of the data varying the two parameters $\beta'$
and the width of the $\rho(1450)$ resonance (fixing its mass to 
$M_{\rho(1450)}= 1.53$ GeV~\cite{Edwards:1999fj}). We included 60
data points  taken from
refs.~\cite{Achasov:2000wy,Akhmetshin:2003ag,Achasov:2012zz} in the
fit and obtain the following values for the parameters 
\be
\beta'= -0.177\pm 0.004,\quad \Gamma_{\rho(1450)}= 0.560\pm
0.024\ \hbox{GeV} 
\en
giving a $\chi^2/N_{dof}=1.51$. These results allow  one to extract
the value of the $\omega\rho\pi$ coupling constant (which will be
useful below) from its relation with the coefficient of the
$\rho(770)$ Breit-Wigner function,
\be
{F_\rho\,g_{\omega\rho\pi} \over 2 m_\rho\, C_\omega}={1\over1+\beta'}
\en
which gives (using $F_\rho=156.44\pm0.67$ MeV)
\be\lbl{gomegarhopi}
g_{\omega\rho\pi}\simeq 13.8\pm0.3\ \hbox{GeV}^{-1}\ .
\en
This value is somewhat smaller than the one obtained in the fits of
ref.~\cite{Akhmetshin:2003ag}
(e.g. $g_{\omega\rho\pi}=16.7\pm0.4\pm0.6$ in fit I). This is because
a) our fit is constrained to reproduce the PDG value of
$\omega\to\gamma\pi^0$ when $q^2\to0$ (see eq.~\rf{Vpiffdef}) and b)
our use of Gounaris-Sakurai functions for the $\rho$, $\rho'$ mesons. 
This indicates that there is a significant model dependent uncertainty
in the determination of $g_{\omega\rho\pi}$ from $e^+
e^-\to\omega\pi$. We account for this by multiplying the error in
eq.~\rf{gomegarhopi} par a factor 10. 
%%%%%%%%%%%%%%figure omega-pi form factor (I)
\begin{figure}
\bc
\includegraphics[width=0.8\linewidth]{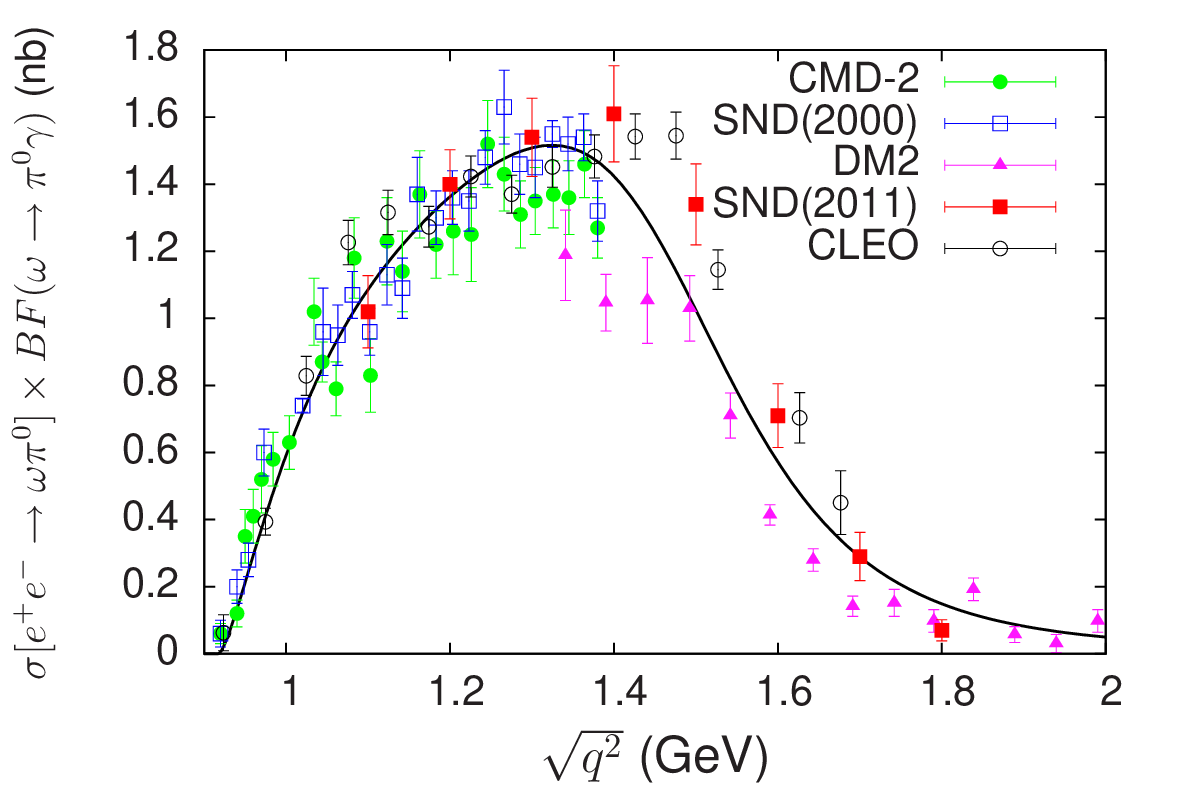}
\caption{\sl Experimental results on the $e^+e^-\to\omega\pi^0$
  cross-section (multiplied by the $\omega\to \pi^0\gamma$ branching
  fraction) fitted to a form factor with two resonances as in
  eq.~\rf{Fomegapi0}. The data shown are from
  refs.~\cite{Akhmetshin:2003ag} (CMD-2),~\cite{Achasov:2000wy}(SND
  (2000)),~\cite{Bisello:1990du} (DM2),~\cite{Achasov:2012zz} (SND
  (2011)) and ~\cite{Edwards:1999fj} (CLEO). }
\lblfig{sigmaVP}
\ec
\end{figure}
%%%%%%%%%%%%%%figure omega-pi form factor (I)

The result of this fit is illustrated on fig.~\fig{sigmaVP}
and on fig.~\fig{ffomegapi} which also shows the energy region below
the $\omega$ peak. One observes a reasonable agreement
between our determined form factor and the data at low energies
$q \lapprox 0.55$ GeV but not with the few data points lying in the
range  $0.60 \le q\le 0.63$ GeV. The pole-like behaviour in this small
region is a puzzle which cannot be explained by theoretical 
models~\cite{Koepp:1974da,Schneider:2012ez,Terschlusen:2012xw}. 
A new $\omega\to e^+ e^-\gamma$ decay experiment is being performed by
the WASA at COSY collaboration~\cite{Schadmand:2012rh} which will
hopefully clarify the situation. 
%%%%%%%%%%%%%%figure omega-pi form factor (II)
\begin{figure}
\bc
\includegraphics[width=0.8\linewidth]{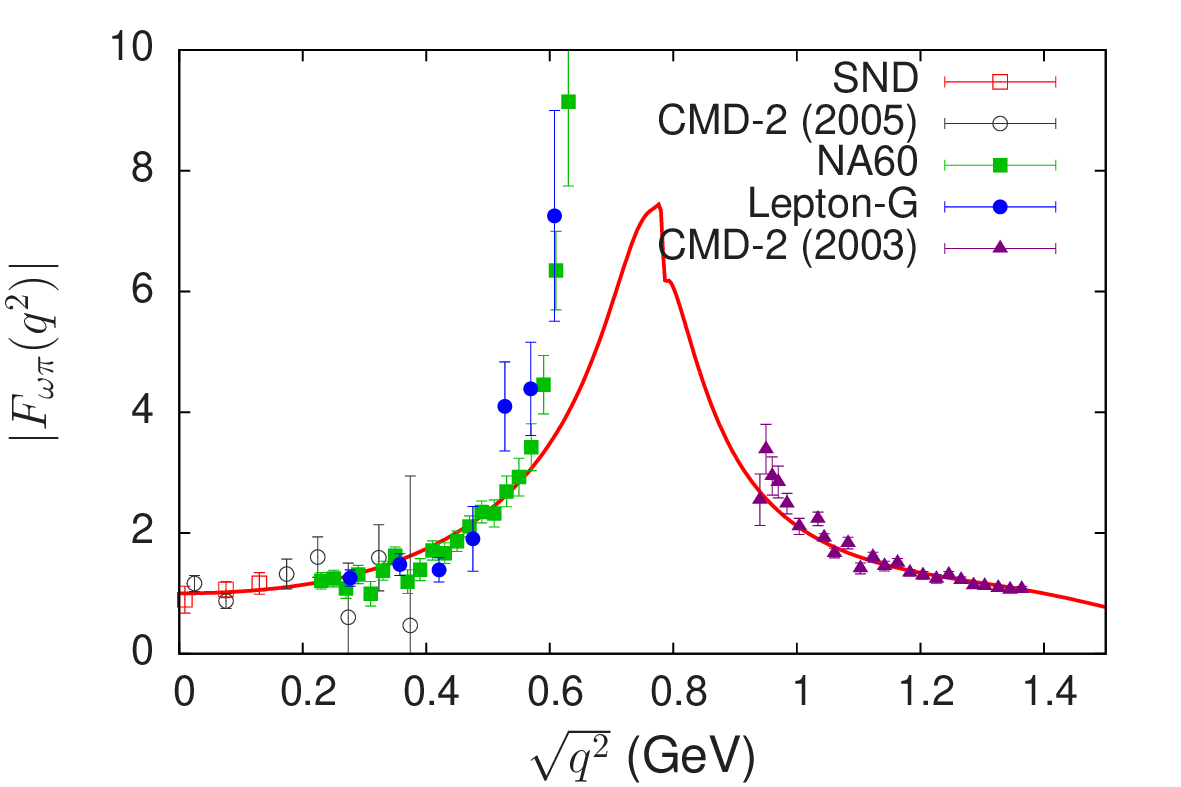}
\caption{\sl  Form factor $\vert F_{\omega\pi}\vert$: the solid curve
  is the fit to the data on $e^+ e^-\to\omega\pi$, it is compared to
  the data on $\omega\to l^+ l^- \pi$ from
  refs.~\cite{Achasov:2008zz}(SND), ~\cite{Akhmetshin:2005vy}
  (CMD-2(2005), ~\cite{Dzhelyadin:1980tj} (lepton-G),
  ~\cite{Arnaldi:2009aa} (NA60).
  }
\lblfig{ffomegapi}
\ec
\end{figure}
%%%%%%%%%%%%%%figure omega-pi form factor (II)

\subsection{ $F_{\rho\pi}(\qdd)$ form factor}
In principle, one could determine the $F_{\rho\pi}$ form factor
similarly to $F_{\omega\pi}$, using experimental inputs from $e^+
e^-\to \rho \pi$ scattering and $\rho\to l^+ l^- \pi$ decay. 
Unfortunately, the width of the $\rho(770)$ resonance is much larger
than that of the $\omega$ and this makes it much more difficult to extract
unambiguously the $e^+ e^-\to \rho \pi$ cross section than it was for
$\omega\pi$. There is also only an upper bound available for the decay
amplitude  $\rho\to e^+ e^- \pi$. We will therefore try to estimate
the $F_{\rho\pi}$ form factor from (hopefully) plausible
phenomenological considerations rather than from actual data.
Let us start by writing, as before, a representation in terms of three
Breit-Wigner functions (we ignore $\rho-\omega$ mixing here), 
\be\lbl{KSforrhopi}
F_{\rho\pi}(\qdd)=\alpha_\omega\, BW_\omega(\qdd)
            + \alpha_\phi \,   BW_\phi(\qdd)
            + \alpha_{\omega'}\,BW_{\omega'}(\qdd) 
\en
with $\alpha_\omega+ \alpha_\phi+ \alpha_{\omega'}=1$ and try to
determine the $\alpha_V$ parameters. The
first one, $\alpha_\omega$, can be related to the $\omega\rho\pi$
coupling constant  
\be\lbl{alphaom}
\alpha_\omega= {F_\omega\, g_{\omega\rho\pi}\over 2 m_\omega C_\rho}
\en
and we can use its value determined above~\rf{gomegarhopi} (with a
rescaled error, accounting for the model dependence), together
with $F_\omega=45.9\pm0.8$ MeV and $C_\rho=0.42\pm 0.02$ from
the second line in eq.~\rf{CtildeV}, which gives
\be
\alpha_\omega= 0.96\pm 0.19\ .
\en
We can write for the second parameter, $\alpha_\phi$, a relation
analogous to eq.~\rf{alphaom} and determine the coupling
$g_{\phi\rho\pi}$ from $\phi\to\rho\pi$. However, only the branching
fraction for $\phi\to3\pi$ is precisely known, so we must content with
a guess: $BF(\phi\to\rho\pi)\simeq 0.8\times BF(\phi\to3\pi)$ which gives
\be
g_{\phi\rho\pi}\simeq -1.09\ \hbox{GeV}^{-1}\ .
\en
The choice of the minus sign can be justified by the consideration of
the $\omega-\phi$ mixing angle $\theta_V$. Indeed, one expects (in a simple
minded quark model picture) the following relation  to hold
\be
{g_{\phi\rho\pi}\over g_{\omega\rho\pi}}= {1-\sqrt2 \tan\theta_V\over
\sqrt2+\tan\theta_V}=\tan(\theta_{id}-\theta_V)
\en
which requires the left-hand side to be negative if $\theta_V >
\theta_{id}=35.26^\circ$. This seems indeed to be the case if one
determines  $\theta_V$ from the vector meson masses, e.g. the
quadratic mass formula gives $\theta_V\simeq 39^\circ$. Our determined
values for the couplings $g_{\phi\rho\pi}$, $g_{\omega\rho\pi}$ gives
a reasonably similar value: $\theta_V\simeq 39.8^\circ$. 
This then leads to the following estimate for the parameter
$\alpha_\phi$, 
\be
\alpha_\phi\simeq-0.101
\en
while the last parameter in the representation~\rf{KSforrhopi} for
$F_{\rho\pi}$ is determined from the normalization condition
$\alpha_{\omega'}=1-\alpha_\omega-\alpha_\phi$. 
Finally, the result for the
form factor is illustrated on fig.~\fig{Frhopi}.
\begin{figure}
\bc
\includegraphics[width=0.8\linewidth]{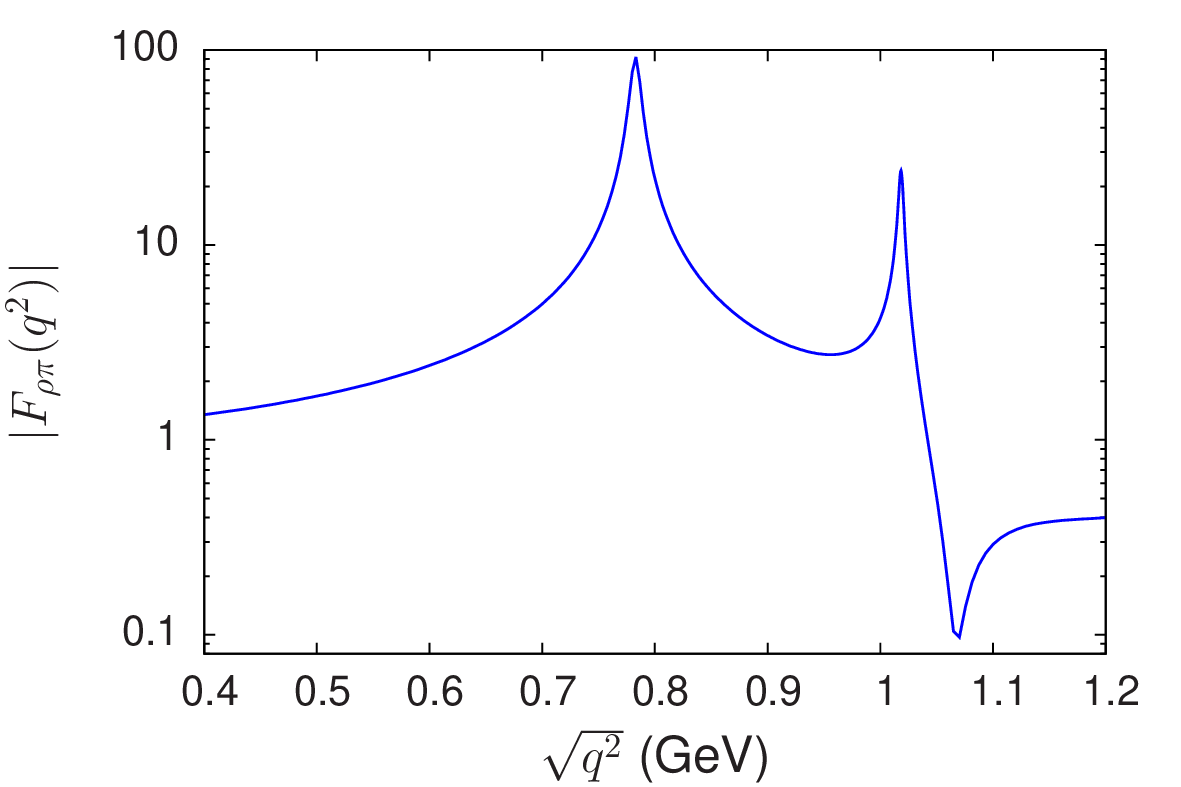}
\caption{\sl Illustration of the modulus of the form factor
  $F_{\rho\pi}$ as modelled from eq.~\rf{KSforrhopi}.}
\lblfig{Frhopi}
\ec
\end{figure}

\bibliography{gstarpaperv2}
\end{document}